\documentclass[superscriptaddress,amsmath,amssymb,aps,prb,twocolumn,nofootinbib,nolongbibliography]{revtex4-2}
\usepackage{graphicx}
\usepackage{graphics}
\usepackage{dcolumn}          
\usepackage{bm}               
\usepackage{upgreek}
\usepackage{booktabs} 				
\usepackage{tabularx}
\usepackage{array}
\usepackage[colorlinks=true,urlcolor=blue,breaklinks=true]{hyperref}
\usepackage[justification=justified,singlelinecheck=false]{caption}
\usepackage{afterpage}
\usepackage{xcolor}

\usepackage{orcidlink}

\usepackage{xr}

\usepackage{siunitx}
\usepackage{stackengine}
\usepackage{svg}
\usepackage{graphicx,subfigure}
\usepackage{placeins}
\usepackage{bibunits}

\usepackage{ragged2e} 

\usepackage{titlesec}
\begin{document}
\title{Aharonov--Bohm and Altshuler--Aronov--Spivak oscillations in the quasi-ballistic regime in phase-pure GaAs/InAs core/shell nanowires}

\author{Farah Basari\'{c}\,\orcidlink{0009-0000-8721-3129}}
\email{f.basaric@fz-juelich.de}
\affiliation{Peter Gr\"unberg Institut (PGI-9), Forschungszentrum J\"ulich, 52425 J\"ulich, Germany}
\affiliation{JARA-Fundamentals of Future Information Technology, J\"ulich-Aachen Research Alliance, Forschungszentrum J\"ulich and RWTH Aachen University, 52425 J\"ulich, Germany}

\author{Vladan Brajovi\'{c}}
\affiliation{Peter Gr\"unberg Institut (PGI-9), Forschungszentrum J\"ulich, 52425 J\"ulich, Germany}
\affiliation{JARA-Fundamentals of Future Information Technology, J\"ulich-Aachen Research Alliance, Forschungszentrum J\"ulich and RWTH Aachen University, 52425 J\"ulich, Germany}

\author{Gerrit Behner\,\orcidlink{0000-0002-7218-3841}}
\affiliation{Peter Gr\"unberg Institut (PGI-9), Forschungszentrum J\"ulich, 52425 J\"ulich, Germany}
\affiliation{JARA-Fundamentals of Future Information Technology, J\"ulich-Aachen Research Alliance, Forschungszentrum J\"ulich and RWTH Aachen University, 52425 J\"ulich, Germany}

\author{Kristof Moors\,\orcidlink{0000-0002-8682-5286}}
\thanks{Present Address: Imec, Kapeldreef 75, 3001 Leuven, Belgium}
\affiliation{Peter Gr\"unberg Institut (PGI-9), Forschungszentrum J\"ulich, 52425 J\"ulich, Germany}
\affiliation{JARA-Fundamentals of Future Information Technology, J\"ulich-Aachen Research Alliance, Forschungszentrum J\"ulich and RWTH Aachen University, 52425 J\"ulich, Germany}

\author{William Schaarman}
\affiliation{Peter Gr\"unberg Institut (PGI-9), Forschungszentrum J\"ulich, 52425 J\"ulich, Germany}
\affiliation{JARA-Fundamentals of Future Information Technology, J\"ulich-Aachen Research Alliance, Forschungszentrum J\"ulich and RWTH Aachen University, 52425 J\"ulich, Germany}

\author{Andrei Manolescu\,\orcidlink{0000-0002-0713-4664}}
\affiliation{Department of Engineering, Reykjavik University, Menntavegur 1, IS-102 Reykjavik, Iceland}

\author{Raghavendra Juluri\,\orcidlink{0000-0002-6500-5983}} \affiliation{Department of Physics, University of Warwick, Coventry CV4 7AL, UK}

\author{Ana M. Sanchez} \affiliation{Department of Physics, University of Warwick, Coventry CV4 7AL, UK}

\author{Jin Hee Bae}
\affiliation{Peter Gr\"unberg Institut (PGI-9), Forschungszentrum J\"ulich, 52425 J\"ulich, Germany}
\affiliation{JARA-Fundamentals of Future Information Technology, J\"ulich-Aachen Research Alliance, Forschungszentrum J\"ulich and RWTH Aachen University, 52425 J\"ulich, Germany}

\author{Hans L\"uth\,\orcidlink{0000-0003-1617-3355}}
\affiliation{Peter Gr\"unberg Institut (PGI-9), Forschungszentrum J\"ulich, 52425 J\"ulich, Germany}
\affiliation{JARA-Fundamentals of Future Information Technology, J\"ulich-Aachen Research Alliance, Forschungszentrum J\"ulich and RWTH Aachen University, 52425 J\"ulich, Germany}

\author{Detlev Gr\"utzmacher\,\orcidlink{0000-0001-6290-9672}}
\affiliation{Peter Gr\"unberg Institut (PGI-9), Forschungszentrum J\"ulich, 52425 J\"ulich, Germany}
\affiliation{Peter Gr\"unberg Institut (PGI-10), Forschungszentrum J\"ulich, 52425 J\"ulich, Germany}
\affiliation{JARA-Fundamentals of Future Information Technology, J\"ulich-Aachen Research Alliance, Forschungszentrum J\"ulich and RWTH Aachen University, 52425 J\"ulich, Germany}

\author{Alexander Pawlis\,\orcidlink{0000-0002-3394-0707}}
\affiliation{Peter Gr\"unberg Institut (PGI-9), Forschungszentrum J\"ulich, 52425 J\"ulich, Germany}
\affiliation{Peter Gr\"unberg Institut (PGI-10), Forschungszentrum J\"ulich, 52425 J\"ulich, Germany}
\affiliation{JARA-Fundamentals of Future Information Technology, J\"ulich-Aachen Research Alliance, Forschungszentrum J\"ulich and RWTH Aachen University, 52425 J\"ulich, Germany}

\author{Thomas Sch\"apers\,\orcidlink{0000-0001-7861-5003}}
\email{th.schaepers@fz-juelich.de}
\affiliation{Peter Gr\"unberg Institut (PGI-9), Forschungszentrum J\"ulich, 52425 J\"ulich, Germany}
\affiliation{JARA-Fundamentals of Future Information Technology, J\"ulich-Aachen Research Alliance, Forschungszentrum J\"ulich and RWTH Aachen University, 52425 J\"ulich, Germany}
\hyphenation{}
\date{\today}

\begin{abstract}
The realization of various qubit systems based on high-quality hybrid superconducting quantum devices, is often achieved using semiconductor nanowires. For such hybrid devices, a good coupling between the superconductor and the conducting states in the semiconductor wire is crucial. GaAs/InAs core/shell nanowires with an insulating core, and a conductive InAs shell fulfill this requirement, since the electronic states are strongly confined near the surface. However, maintaining a good crystal quality in the conducting shell is a challenge for this type of nanowire. In this work, we present phase-pure zincblende GaAs/InAs core/shell nanowires and analyze their low-temperature magnetotransport properties. We observe pronounced magnetic flux quantum periodic oscillations, which can be attributed to a combination of Aharonov--Bohm and Altshuler--Aronov--Spivak oscillations. From the gate and temperature dependence of the conductance oscillations, as well as from supporting theoretical transport calculations, we conclude that the conducting states in the shell are in the quasi-ballistic transport regime, with few scattering centers, but nevertheless leading to an Altshuler--Aronov--Spivak correction that dominates at small magnetic field strengths. Our results demonstrate that phase-pure zincblende GaAs/InAs core/shell nanowires represent a very promising alternative semiconductor nanowire-based platform for hybrid quantum devices.
\end{abstract}
\maketitle

\section{Introduction}
\begin{bibunit}[]
Semiconductor nanowires are considered to be a versatile building block for various applications in nanoelectronics and quantum devices \cite{Badawy2024}. In particular, when combined with superconductors, they provide a flexible platform for various applications in classical and topological quantum computing \cite{deLange2015,Larsen2015,Tosi2019,Metzger2021,Zellekens2022,Yazdani2023}. In many cases bulk InAs nanowires are used, since the Fermi level pinning at the surface naturally leads to an accumulation layer at the interface in the semiconductor \cite{Wirths2011}. This ensures conductive channels even at low temperatures. The accumulation layer also provides good coupling to metallic contacts, which is particularly important in the case of superconducting electrodes. Even better coupling is expected when the electronic state in the nanowire is confined close to the interface with the superconductor by using a semiconductor heterostructure in a core/shell nanowire.

In a core/shell nanowire, the heterointerface between the high bandgap core semiconductor and the low bandgap shell material, creating a band offset, forms a radial quantum well that confines the electron wave function close to the outer radius. This results in a tubular conductor \cite{Tserkovnyak2006,Rosdahl2014,Ferrari2008,Manolescu2016}. Applying an axial magnetic field leads to magnetic flux quantum $h/e$-periodic Aharonov--Bohm-type (AB) oscillations in the magnetoconductance \cite{Aharonov1959}, where $e$ is the electron charge and $h$ is Planck's constant. Aharonov--Bohm-type oscillations have been observed in GaAs/InAs \cite{Guel2014,Bloemers2013,Haas2016,Haas2017}, GaAs/InSb \cite{Zellekens2020}, as well as In$_2$O$_3$/InO$_x$ core/shell nanowires \cite{Jung2008}. The tubular quantum states in semiconductor core/shell nanowires have certain similarities to the tubular topological protected surface states in a topological insulator nanowire \cite{Peng2010,Cho2015,Arango2016}. For this type of nanowire, AB oscillations in the quasi--ballistic regime have recently been observed. A possible reason for reaching the quasi-ballistic regime in this case is that spin-momentum locking in topological insulators leads to reduced scattering \cite{Ziegler2018,Rosenbach2022}. On the other hand, Altshuler--Aronov--Spivak (AAS) oscillations with a period of $h/2e$ resulting from interference of time-reversed paths have been observed here as well \cite{Kim2020}, indicating the presence of scattering.  

Recently, we have succeeded in growing GaAs nanowires with a phase-pure crystal structure by dynamically controlled growth using molecular beam epitaxy (MBE) \cite{Jansen2020}. In this study, we use this approach to grow zincblende phase-pure GaAs/InAs core/shell nanowires and investigate their transport properties. Due to the phase purity in the InAs shell, we expect significantly reduced scattering in electronic transport. In this regime of reduced scattering we investigate magnetotransport oscillations that are known to appear in both (quasi-)ballistic and diffusive regimes. Gate-dependent low temperature transport measurements allow us to disentangle different contributions to the oscillatory behavior of the magnetoconductance. The relevant transport regime is determined by temperature dependent measurements. We interpret the experimental results by comparing them with transport calculations based on linear response theory (Kubo formalism) and the Landauer formalism (quantum transport simulations using KWANT) \cite{Groth2014}.

\section{Experimental details}

The GaAs/InAs core/shell nanowires are grown by MBE via self-catalyzed vapor-liquid-solid technique on pre-structured substrates. To obtain a phase-pure crystal structure, the catalyst droplet on top of the growing nanowires is dynamically controlled during the growth, following the approach reported in Jansen \textit{et al.} \cite{Jansen2020}. In our case, we achieved a contact angle of the Ga catalyst droplet above $125^\circ$ resulting in a zincblende (ZB) crystal structure.  We used Si(111) substrates covered with about 20\,nm of thermally deposited $\mathrm{SiO}_{2}$ containing hole arrays with varying diameters of 40, 60, and 80\,nm and pitches with varying pinhole sizes of 0.5, 1, 2, and 4\,$\upmu$m, prepared by electron beam lithography and subsequent dry and wet etching. The GaAs core is grown by applying an As flux with a beam equivalent pressure (BEP) of $5\times 10^{-6}$ mbar for 90\,min at about $610^\circ$C, while the Ga flux is dynamically decreased by 40\,\% from the starting value of $1.5\times 10^{-7}$\,mbar. Subsequently, the InAs shell growth is performed at a substrate temperature of $450^\circ$C comprising In and As fluxes with BEPs of $1.95\times 10^{-7}$ mbar and $5\times 10^{-6}$ mbar, respectively. 

\begin{figure}[!htbp]
    \centering 
    \includegraphics[width=0.46\textwidth]{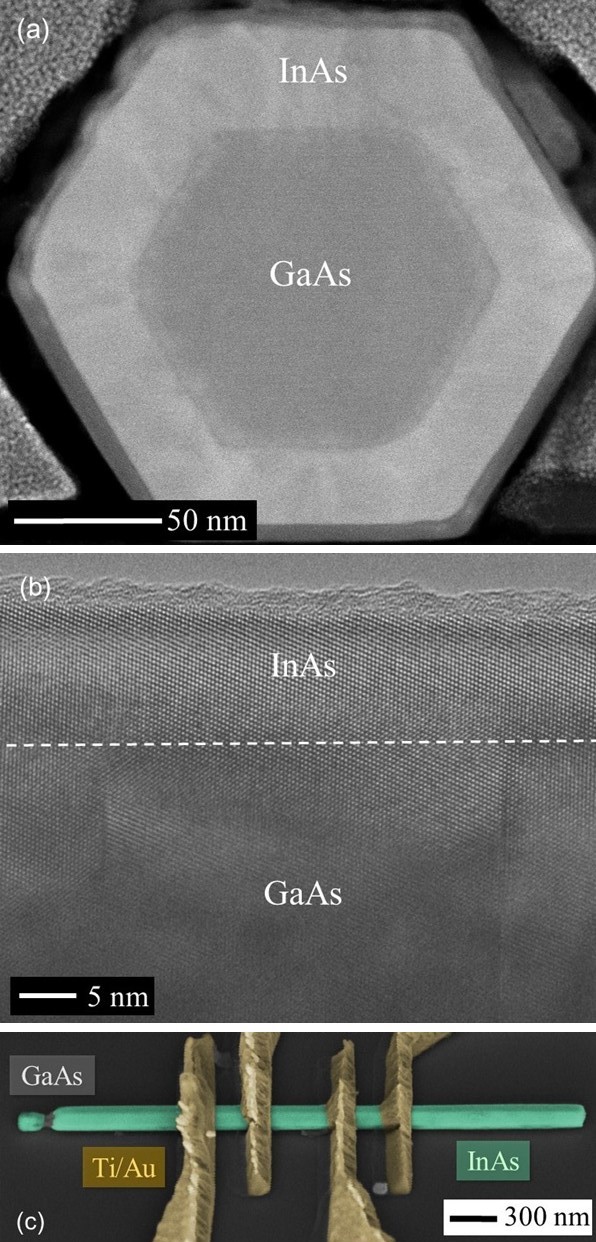}  
    \caption{\justifying(a) ADF-STEM image of the GaAs/InAs core/shell nanowire cross-section of sample B along the growth direction. (b) High-resolution TEM image of the middle part of a core/shell nanowire. (c) False color SEM image of a contacted GaAs/InAs nanowire, corresponding to sample A.} 
    \captionsetup{position=bottom}
    \label{fig:Figure-SEM}
\end{figure}

In this work, measurements are discussed on three samples, originating from two separate growth runs. In case of samples A and B the total growth time of 25\,min for the InAs shell resulted in a smaller shell thickness $t_\mathrm{shell}$ compared to sample C with a shell growth time of 40\,min (cf. Table~\ref{tab:Geometry&General}). Figure~\ref{fig:Figure-SEM} (a) corresponds to a low magnification annular dark field scanning transmission electron image (ADF-STEM) of the nanowire cross-section, i.e. along the $<$111$>$ direction.  Figure~\ref{fig:Figure-SEM} (b) shows a high-resolution TEM image along the $<$110$>$ direction. This type of images confirmed that ZB crystal structure is the preferred polytype both in the core and the shell. Detailed information on the growth procedure and the crystal structure is given in Supporting Information.

For the low-temperature transport experiments, the GaAs/InAs core/shell nanowires are transferred by a scanning electron microscope (SEM)-based micro-manipulator to a pre-patterned Si(100) substrate covered with $150\,\mathrm{nm}$ of $\mathrm{SiO_{2}}$. Employing the precise nanowire transferring technique, as well as the electron beam lithography step and Ar$^\mathrm{+}$ cleaning, the metallization of Ti/Au contact fingers was realized. An example of a measured device is given in Fig.~\ref{fig:Figure-SEM} (c). As previously mentioned, the measurement results of samples A and C are described here, with additional data from a second nanowire of the first growth run (sample B) summarized in the Supporting Information. The corresponding geometric dimensions of these nanowires are given in Table \ref{tab:Geometry&General}. 
\begin{table}[h]
    \centering
    \begin{tabular}{c|c|c|c|c|c}
         \hline
 Sample & $r_{\mathrm{core}}\,(\mathrm{nm})$ & $t_{\mathrm{shell}}\,$(nm) & $r_{\mathrm{tot}}\,$(nm) & $L_{\mathrm{c}}\,$(nm) & $l_{\varphi}\,(\mathrm{\upmu m})$\\ 
 \hline
 A & 51 & 28 & 79 & 400 & 3.3\\
 B & 58 & 27 & 85 & 400 & 2.7 \\
 C & 56 & 39 & 95 & 450 & 4.6\\ 
 \hline
    \end{tabular}
    \caption{\justifying Geometric dimensions and phase-coherence length of samples A to C, where $r_{\mathrm{core}}$ denotes the GaAs core radius, $t_{\mathrm{shell}}$ the InAs shell thickness, $r_{\mathrm{tot}}$ the total nanowire radius, $L_{\mathrm{c}}$ the inner contact separation, and $l_{\varphi}$ the phase-coherence length obtained at 1.5\,K.}
    \label{tab:Geometry&General}
\end{table}

The measurements were carried out in a variable temperature insert cryostat, with a base temperature of 1.5\,K. Using a four-terminal measurement configuration and a standard lock-in setup, as well as applying an axial magnetic field, the devices were biased with a current of $20\,\mathrm{nA}$ between the two outer contacts, whereas the voltage drop was measured between the two inner ones. The inner contact separation $L_c$ for the investigated samples is given in Table \ref{tab:Geometry&General}. The highly-doped Si substrate was used as a global backgate with the $\mathrm{SiO_{2}}$ layer used as a gate dielectric. 

\section{Results and Discussion}

\subsubsection{Gate dependence}

\begin{figure*}[!htbp]
    \centering 
    \includegraphics[width=\textwidth]{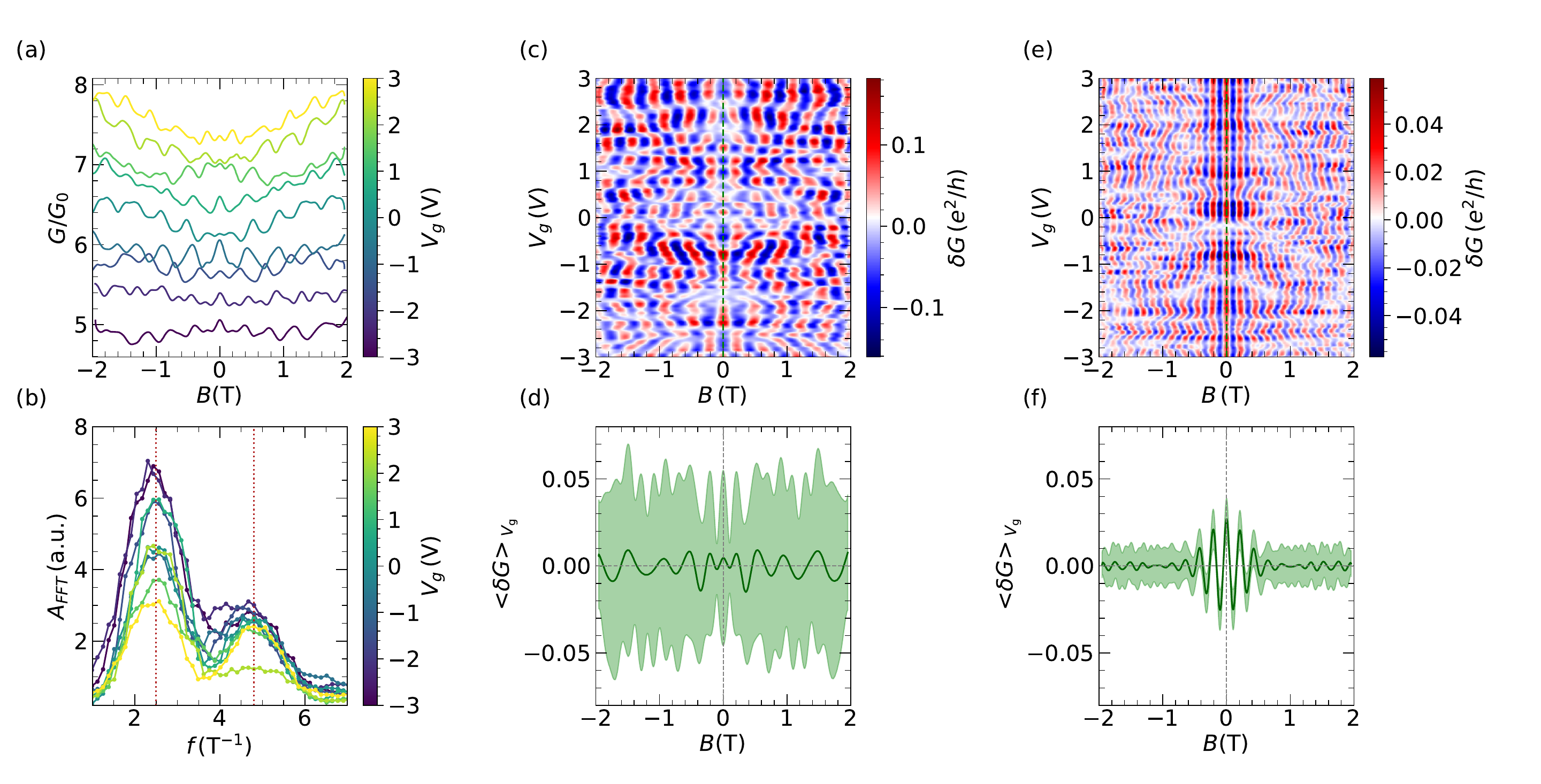}  
    \caption{\justifying(a) Normalized conductance $G/G_0$ of sample A, with $G_0=e^2/h$, as a function of the magnetic field, displayed for some values of applied gate voltage at 1.5\,K, and oscillating with a period of $0.4$\,T. (b) Fast Fourier transform of the magnetoconductance oscillations shown in (a) featuring a two peak structure, corresponding to $h/e$-, and $h/2e$-periodic oscillation frequencies at 2.5 and 4.8 $\mathrm{T^{-1}}$, respectively. (c) Color map of filtered out $h/e$-periodic conductance contribution as a function of $B$ and $V_\mathrm{g}$. It reveals a non-rigid phase with varying $V_\mathrm{g}$ over the entire magnetic field range. (d) Oscillating $h/e$ conductance contribution averaged over the gate voltage range $<\!\delta G\!>_{V_\mathrm{g}}$ The light green area shows the standard deviation. (e) Analogous color map analysis for $h/2e$-periodic contribution to the one shown in (c). It shows phase-rigidity at a small magnetic field range between $-0.3$ and $0.3$\,T. (f) Averaged $h/2e$-periodic conductance contribution and standard deviation as in case of (d).
    }
    \captionsetup{position=bottom}
    \label{fig:Figure-Gate}
\end{figure*}

First, we focus on the gate-dependent magnetotransport properties of sample A (the corresponding measurements and analysis of samples B and C are given in Supporting Information). The measurements were conducted at 1.5\,K in an axial magnetic field. A gate voltage $V_\mathrm{g}$ ranging between $-3$ and 3\,V, with a stepping of $0.05$V was applied to the global backgate. As shown in Fig.~\ref{fig:Figure-Gate} (a), the normalized magnetoconductance $G/G_0$, with $G_0=e^2/h$, reveals regular AB-type oscillations over the entire gate voltage range. 

From the oscillation period $\Delta B$, the effective area $S$ enclosed by the closed-loop wave function can be determined by the relation $\Phi_0=\Delta B \cdot S$, where $\Phi_0=h/e$ is the magnetic flux quantum. Assuming a hexagonal nanowire cross-section, the area is given by $S\,=\,r^{2}(3\sqrt{3}/2)$, where $r$ is the the circumradius of the hexagon that is enclosed by the closed-loop wavefunction. For the extracted period of $\Delta B=0.4$\,T, we get a radius of $r=56\,\mathrm{nm}$, confirming that the wave function is located within the InAs shell.

Aharonov--Bohm oscillations are related to a periodic modulation of the average occupation of the phase-coherent conducting states with different quantized transverse momenta enveloping the core \cite{Guel2014}. By shifting the Fermi level position with the gate voltage, the average occupation is either forming a local maximum or minimum at $B=0$, giving rise to phase jumps as a function of the gate voltage between $\theta = 0$ and $\pi$ in the magnetoconductance oscillation pattern: $\delta G(\phi) \propto \cos(\phi + \theta(V_\mathrm{g}))$, with $\phi=\Phi/\Phi_0$ the normalized magnetic flux and $\theta(V_\mathrm{g})$ the gate voltage-dependent phase of the oscillation pattern. Note that the average occupation and the corresponding phase in the magnetoconductance oscillation pattern can vary over different parts of the sample under the change of local electron density in an environment of randomly distributed scattering centers within the InAs shell.

To further analyze the experimental data, a fast Fourier transform (FFT) of the magnetoconductance oscillations is performed. Before applying the FFT, the slowly varying background $\overline{G}$ in the experimental data was subtracted from the measured conductance, resulting in an oscillation signal $\delta G=G-\overline{G}$. As can be seen in Fig.~\ref{fig:Figure-Gate} (b), the FFT spectrum obtained from measurements at different gate voltages systematically shows a two peak structure. The, first peak can be assigned to the AB oscillations with a period of $h/e$ while the second, smaller, peak corresponds a period of $h/2e$ . The latter can originate from both AAS-type conductance oscillations or higher-harmonic contributions of the AB oscillations. The AAS oscillations result from interference of time–reversed paths in a tubular conductor \cite{Altshuler1981,Sharvin81}. 

To gain a deeper insight into the physical origin of the two quantum transport contributions, we analyzed and compared the phase rigidity of the $h/e$- and $h/2e$-periodic oscillations. For this purpose, the inverse Fourier transform of $\delta G$ was evaluated. Furthermore, we isolated specific frequency windows in the Fourier spectrum corresponding to the peaks belonging to periods of $h/e$ and $h/2e$, respectively. Such an analysis is shown in Fig.~\ref{fig:Figure-Gate} (c) for the $h/e$-periodic oscillations, revealing a complex oscillation pattern with non-rigid phase as $V_\mathrm{g}$ changes. Since the $h/e$ contribution dominates the oscillation pattern, the phase jumps along $B=0$ follow the phase jumps in the raw data. As a consequence of the random phase shifts with $V_\mathrm{g}$, the conductance averaged over the gate voltage $<\!\delta G\!>_{V_\mathrm{g}}$ basically cancels out while at the same time the standard deviation is very large, as can be seen in Fig.~\ref{fig:Figure-Gate} (d). The $h/2e$ oscillations, corresponding to the second peak in the FFT spectrum  are not clearly visible in the raw data, i.e. in Fig.~\ref{fig:Figure-Gate} (a), due to the dominance of $h/e$-periodic oscillations. However, these oscillations are resolved in the inverse FFT spectrum shown in Fig.~\ref{fig:Figure-Gate} (e). Interestingly, in a field range between $-0.3$ and $0.3$\,T, the $h/2e$-periodic oscillations are phase-rigid with a conductance maximum at zero field. Beyond this range the phase rigidity is weakened and eventually lost. The extent of the phase stability can also be deduced from Fig.~\ref{fig:Figure-Gate} (f), which shows the $h/2e$ conductance oscillations averaged over the entire gate voltage range $<\!\delta G\!>_{V_\mathrm{g}}$. It can be seen that $<\!\delta G\!>_{V_\mathrm{g}}$ has its maximum value at $B=0$ and then decreases continuously with increasing field, disappearing completely at about $\pm 0.6$\,T. The small standard deviation in this range underlines the claim of phase stability, as can be seen in Fig.~\ref{fig:Figure-Gate} (f). We attribute the phase-rigid part to AAS oscillations. In terms of their rigidity while changing $V_\mathrm{g}$, AAS oscillations are robust to averaging since they originate from interference of time-reversed paths in our tubular InAs shell, as already resolved for polymorphic GaAs/InAs nanowires \cite{Guel2014}. The conductance maximum at zero field indicates the presence of spin-orbit coupling due to the weak antilocalization effect \cite{Aronov1987}. Such a conductance maximum has also been observed in bulk InAs nanowires \cite{Estevez2010}. 

The phase-rigidity of the $h/2e$-periodic oscillation amplitude up to about $\pm 0.3$\,T (cf.  Fig.~\ref{fig:Figure-Gate} (e)) can be explained by the finite thickness of the InAs shell and the corresponding loss of phase matching along the inner and outer radius when an axial magnetic field is applied. Using the approach outlined in Mur \textit{et al.} \cite{Mur2008}, we obtained a limit of 0.17\,T for our shell cross-section. This value is in relatively good agreement with the experimentally observed one. Beyond this magnetic field limit, the phase relation of the time-reversed paths is randomized \cite{SLRen2015,CPUmbach}. At larger field strengths, the randomized phase relation suppresses the amplitude of $h/2e$ oscillations, which can include higher-harmonic contributions of the $h/e$ AB oscillations as well. As a consequence phase rigidity is lost.

\begin{figure*}[!htbp]
    \centering
    \includegraphics[width=\textwidth]{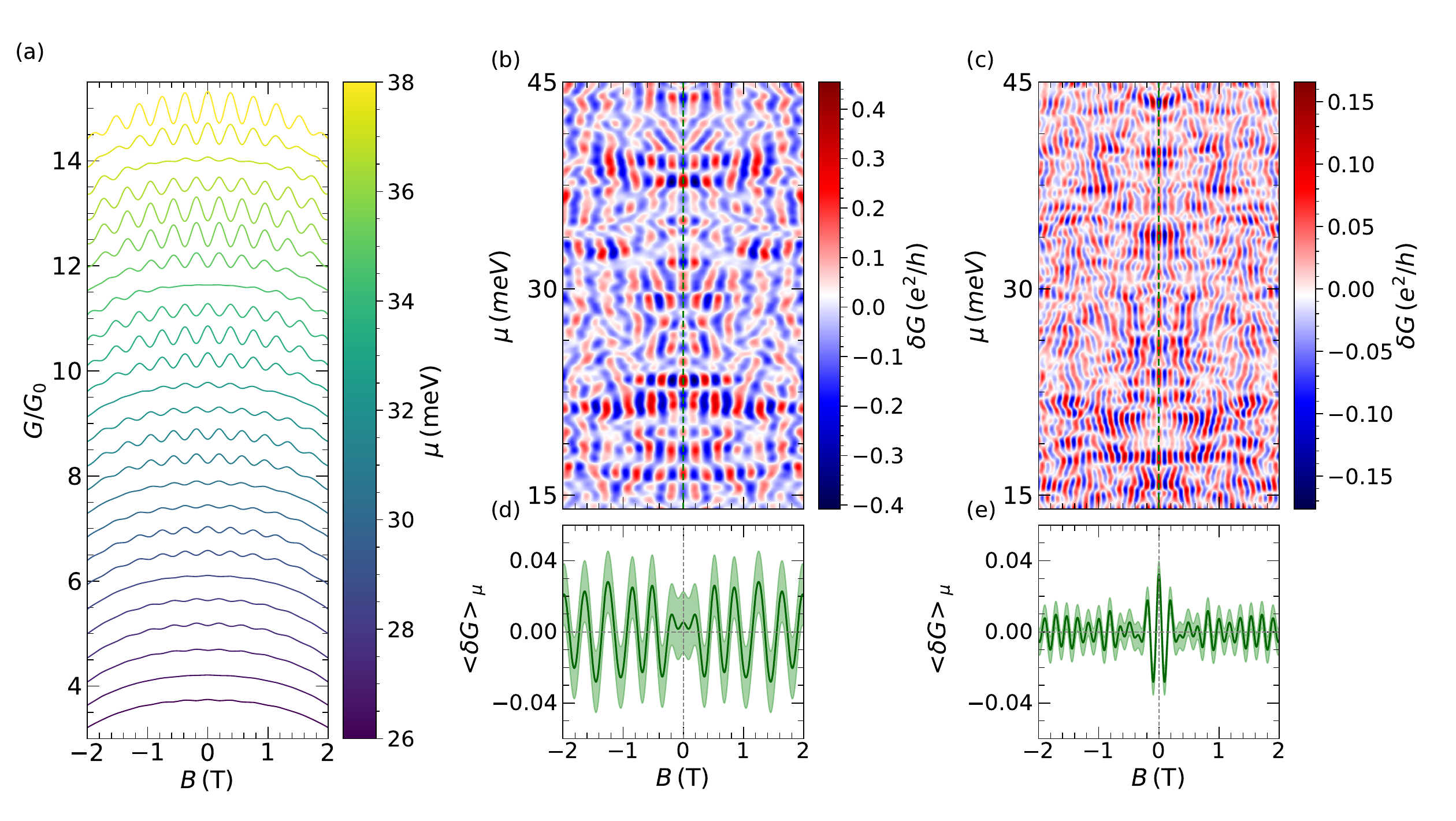}
    \caption{\justifying (a) Normalized magnetoconductance using the Kubo formalism for various values of the chemical potential for a disorder potential of 0.18\,meV at 1.5\,K. (b) Filtered $h/e$ conductance contribution as a function of $B$ and chemical potential $\mu$ of a 4\,$\upmu$m long wire with a disorder potential of 5\,meV at 1.5\,K, calculated using KWANT package. The geometrical dimensions correspond to sample A. (c) Corresponding filtered $h/2e$-periodic oscillations. (d) Averaged oscillation amplitude $<\!\delta G\!>_{\mu}$ over four different disorder potential configuration sets shown by a dark green line, and corresponding standard deviation shown in green. (e) Corresponding averaged oscillation amplitude $<\!\delta G\!>_{\mu}$ and standard deviation for the $h/2e$-periodic contribution.
    }
    \captionsetup{position=bottom}
    \label{fig:Figure-Calc}
\end{figure*}

Several features of the experimentally observed oscillations can be reproduced by transport calculations based on linear response theory using the Kubo formalism \cite{Kubo1957}. This is demonstrated in Fig.~\ref{fig:Figure-Calc} (a), which shows the corresponding calculations of the magnetoconductance. The calculations were performed for the cross-sectional dimensions of sample A for different chemical potentials in the expected range of GaAs/InAs nanowires of similar dimensions \cite{Guel2014,Haas2016}, i.e. for our simulations we assumed a range between 26 and 38\,meV. The Zeeman effect was included with a g-factor of $-14.9$ \cite{Winkler2003}. The general oscillation pattern as well as the phase jumps as a function of chemical potential are well reproduced. Furthermore, by changing the chemical potential, the phase of the oscillations switches. This is in agreement with the experimental results, since applying a gate voltage results in a shift of the chemical potential. However, Kubo formalism assumes an infinite wire length with included disorder. In this approach, the disorder is modeled as a system of uniformly distributed point scatterers, within the self-consistent Born approximation, as summarized in the Supporting Information. For the applied a disorder potential of 0.18\,meV, we obtain oscillation amplitudes comparable to the experimental ones, although the total conductance is somewhat larger. Note, that higher order conductance corrections, represented by vertex or Cooperon diagrams, which are important to capture quantum interference effects, and in particular AAS oscillations, are not contained in our calculations. Thus, possible second harmonic features solely originate from the AB effect, due to discrete variation of the number of transverse channels when the magnetic field increases, the finite thickness of the tubular shell, or its non-circular (hexagonal) cross section. 

In order to include the AAS effect in the modelling, tight binding quantum transport simulations were performed using the software package KWANT \cite{Groth2014}. In fact, by this, a regime can be explored, where the transport is governed by a finite number of scattering centers, being the appropriate scenario for our phase-pure nanowires. For the tight-binding description of the core/shell nanowire, a hexagonal cross-section corresponding to sample A and a wire length of 4\,$\upmu$m was considered. For a direct comparison with the experimental values shown in Figs.~\ref{fig:Figure-Gate} (c) and (d), the filtered first and second harmonics of the simulated conductance are plotted in Figs.~\ref{fig:Figure-Calc} (b) and (c), as a function of magnetic field $B$ and chemical potential $\mu$ at a temperature of 1.5\,K. The general features of the experimentally obtained results are reproduced by the simulations, i.e. the $h/e$ oscillations are non-phase-rigid, while for the $h/2e$ oscillations, the phase rigidity is largely preserved around zero magnetic field. Furthermore, due to the implemented spin-orbit coupling in the simulations, a maximum of $\delta G$ is found at zero field. To further support our reasoning, three additional simulations with different disorder configurations were performed. The results of each of these simulations are presented in the Supporting Information. Figures~\ref{fig:Figure-Calc} (d) and (e) show the conductance oscillations $<\!\delta G\!>_{\mu}$ averaged over the chemical potential as a function of magnetic field of the $h/e$ and $h/2e$ periodic contribution, respectively. Each set is averaged over the chemical potential $\mu$, and presented with the standard deviation. The average over the $h/e$ oscillations shows that the signal is significantly damped due to the phase changes with the chemical potential, and the standard deviation extends over a relatively large range. This indicates that the average phase for the four configurations differs to a large degree. In contrast, we find a different behavior for the averaged $h/2e$-periodic conductance contribution, as shown in Fig.~\ref{fig:Figure-Calc} (e). A clear peak is observed at $B=0$, consistent with phase-rigid AAS oscillations and the presence of spin-orbit coupling. Although a peak at $B=0$ is also observed for the $h/e$ oscillations, it does not correspond to robust phase rigidity. This can be seen from the large standard deviation of the $<\!\delta G\!>_{\mu}$ pattern and the oscillation amplitude being much lower as compared to the typical oscillation patterns for fixed chemical potential. Interestingly, phase rigidity (associated with a maximum at zero field) was obtained only by considering the full length of 4\,$\upmu$m of the nanowire, whereas by assuming a voltage probe separation of 400\,nm, i.e. the separation of the inner contacts, only a random phase is found. We attribute this behavior to the fact that in the case of the 400\,nm wire length, the number of scattering centers is too small to result in a sufficient number of time-reversed paths. It effectively means that the entire wire length is probed. Such a non-local transport effect is more commonly observed in mesoscopic conductors, which allow an extended region of the wire defined by $l_\varphi$ to be probed regardless of the relatively small distances of the voltage probes \cite{Umbach1987}. Summarizing these results, it appears that AAS oscillations show up in our measurements and simulations, which are usually attributed to  higher-order scattering processes, despite the fact that the wire probably has a small number of scattering centers due to the phase purity of the crystal. To further clarify this issue and to find out about the relevant transport regime, temperature-dependent measurements were performed.

\subsubsection{Temperature dependence}

\begin{figure*}[!htbp]
    \centering
    \includegraphics[width=\textwidth]{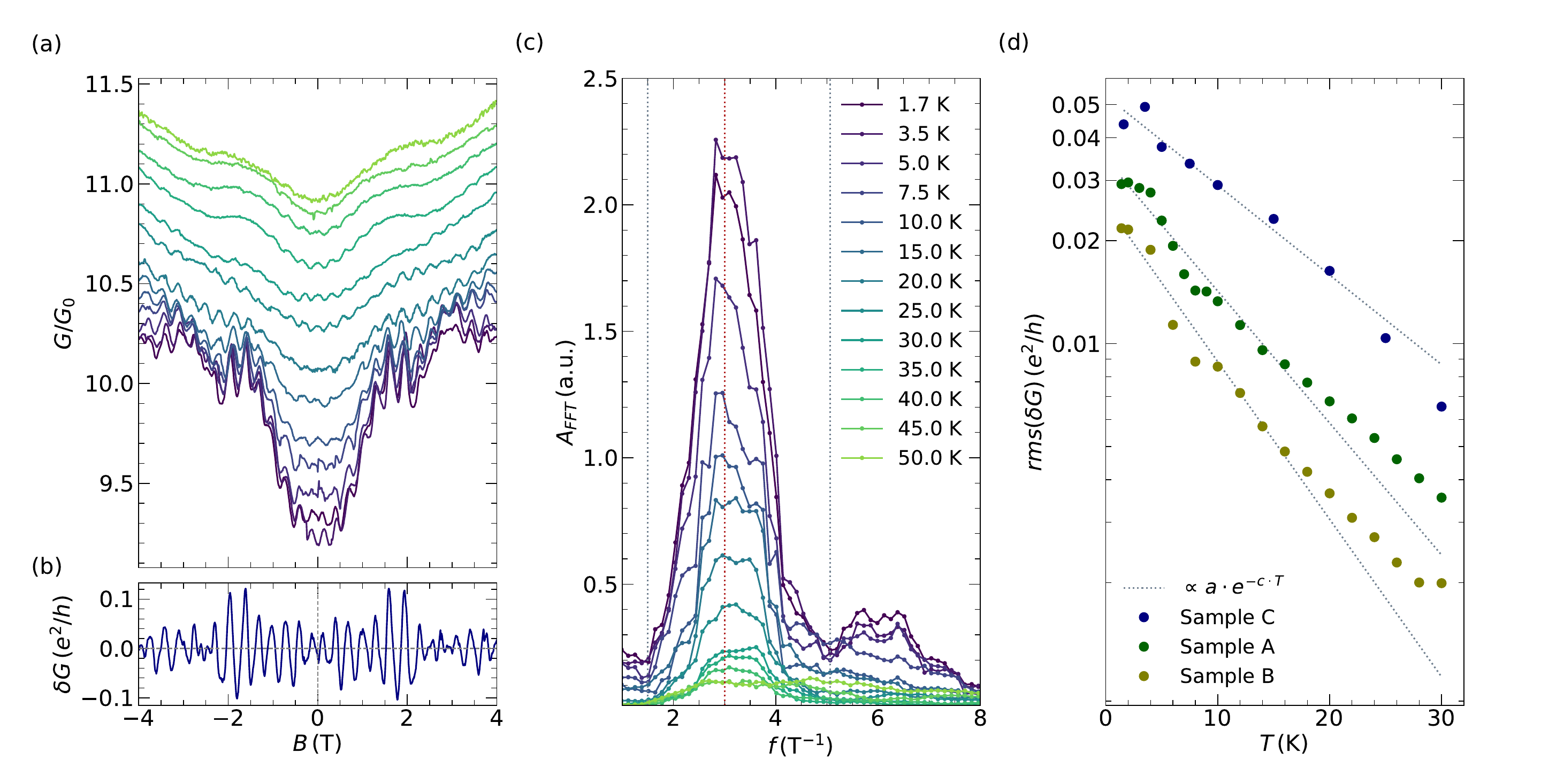}
    \caption{\justifying(a) Temperature-dependent normalized conductance $G/G_0$ under an in-plane magnetic field for sample C. (b) Extracted oscillation pattern after subtracting the slowly varying background signal as a function of magnetic field at 1.7\,K. (c) Fast Fourier transform of the temperature dependent measurement shown in (a). The red line marks the peak central position, associated with radius of enclosed wave-function within the InAs shell. The gray lines indicate the bounds of the shell. (d) Root-mean-square (rms) of the Aharonov--Bohm oscillation amplitude as a function of temperature for the three measured samples and corresponding quasi-ballistic exponential fit. The raw data for samples A and B is offset for clarity by 0.01 $e^2/h$.} 
    \captionsetup{position=bottom}
    \label{fig:Figure-Temp}
\end{figure*}

In this section, temperature-dependent magnetoconductance measurements for sample C are presented, whereas detailed corresponding descriptions for samples A and B are given in Supporting Information. Figure~\ref{fig:Figure-Temp} (a) shows the normalized magnetoconductance $G/G_{0}$ as the temperature was varied from 1.7 to 50.0\,K. Because of the different cross-section of sample C compared to A, here we find magnetoconductance oscillations, with a period of $\Delta B\,=\,0.35\,\mathrm{T}$. The oscillations are superimposed on slower varying conductance modulations, corresponding to universal conductance fluctuations \cite{Lee1987}. With increasing temperature, the number of inelastic scattering events increases, resulting in a reduction of the phase coherence length $l_\varphi$. This is manifested in a gradual decrease of the oscillation amplitude with rising temperature until they are completely suppressed at around 45.0\,K. Figure~\ref{fig:Figure-Temp} (b) shows the oscillation pattern at 1.7\,K after subtracting the slowly varying background signal. The beating pattern found here is attributed to the effect of Zeeman splitting, which causes spin-dependent shifts in the flux parabolas of the energy spectrum \cite{Rosdahl2014}.

The periodic features in the magnetoconductance are analyzed in detail by applying an FFT, as presented in Fig.~\ref{fig:Figure-Temp} (c). The FFT shows a clear peak at a frequency $f$ of about 3.0\,T$^{-1}$, corresponding to the extracted oscillation period given above. The frequency peak indicated by the red dashed line lies within the frequency limits given by the area defined by the inner and outer bounds of the InAs shell. Note that a peak belonging to the second harmonic frequency at about 6.0\,T$^{-1}$ is also observed. With increasing temperature, the peak amplitudes decreases, corresponding to a reduction in the oscillation amplitude. At about 45.0\,K the peak is completely suppressed. 

An essential insight into the phase-coherent transport of phase-pure core/shell GaAs/InAs nanowires under an in-plane magnetic field is provided by the temperature dependence of the $h/e$-periodic oscillation amplitude \cite{Haas2017}. To resolve it, the slowly varying background was first subtracted from the measured conductance, resulting in an oscillation signal $\delta G$, where the average oscillation amplitude is determined by the root mean square $\mathrm{rms}(\delta G)$. Such analysis was carried out for all three samples, and Fig.~\ref{fig:Figure-Temp} (d) shows the best fit for temperature dependence of $\mathrm{rms}(\delta G)$. From this, it is possible to determine the phase coherence length $l_{\varphi}$, considering an exponential decay $\mathrm{rms}(\delta G)\,\propto\, \mathrm{exp}(-2\pi r/l_{\varphi})$ \cite{Webb86}. For all samples, the resulting dependence can be fitted well to the data with $l_{\varphi}\,\sim \,T^{-1}$. The exponential decay of the $h/e$-periodic oscillation amplitude suggests a quasi-ballistic mesoscopic transport regime \cite{Seelig2001}. Even though the experimental data at higher temperature seems to deviate from this fit, it still is the best representation for all three of our samples. Here, we compared between models that describe either the diffusive or quasi-ballistic transport regime, with or without thermal broadening. This is further displayed and discussed in the Supporting Information. The obtained dependence is in contrast to previous measurements on core/shell GaAs/InAs nanowires, where a dependence $l_\varphi \sim T^{-p}$ with $p$ of about 0.6 was extracted \cite{Haas2017}, which is close to the diffusive case \cite{Ludwig2004}. We attribute the different behavior to the improved crystal quality of the present nanowires. From the fit, we determined $l_{\varphi}$ at base temperature for each sample (cf. Table \ref{tab:Geometry&General}). As expected for the ballistic case, $l_\varphi$ exceeds the length of the ring-like tubular conductor arm and is found to be significantly larger than previously reported for GaAs/InAs nanowires \cite{Haas2017}. Additionally, it is in good agreement with the proposed superior transport properties of phase-pure core/shell nanowires. A quasi-ballistic transport regime has also been observed in topological nanowire structures, where spin-momentum locking is assumed to be responsible for the reduced scattering \cite{Dufouleur2013,Ziegler2018,Rosenbach2022}. 

\section{Conclusions}

In conclusion, phase-coherent transport is studied in phase-pure zincblende GaAs/InAs core/shell nanowires. Pronounced Aharonov--Bohm oscillations with a period of $h/e$ are observed. These oscillations arise from the presence of coherent closed-loop states in the InAs shell, which are periodically modulated when a magnetic flux is threaded through the wire cross-section. In contrast to previous studies, the temperature dependence of the oscillation amplitude indicates that the transport takes place in the quasi-ballistic regime. When the electron concentration is varied by a back-gate voltage, the phase of the $h/e$-periodic oscillations are found to be non-phase-rigid, which is attributed to the lack of time-reversal symmetry of this interference process. Interestingly, the $h/2e$ oscillation component contains a phase-rigid region around zero magnetic fields. This can be explained in the framework of Althshuler--Aronov--Spivak oscillations, where time-reversal symmetry is preserved. It is noteworthy that these oscillations are observed in the quasi-ballistic regime and not, as usual, in the diffusive regime. At larger magnetic fields, only the second harmonic of the Aharonov--Bohm oscillations remains. The findings on the presence or absence of phase rigidity of the $h/2e$-periodic oscillations are confirmed by quantum transport simulations, where a relatively small number of scattering centers, represented by a disorder potential with a correlation length of 50\,nm, are found to conform to the quasi-ballistic regime and still give rise to phase-rigid Altshuler--Aronov--Spivak oscillations near zero field. The phenomena and the transport regime observed here are very similar to experiments on topological insulator nanoribbons, despite the absence of spin-momentum locking in our case. The quasi-ballistic transport behavior in our nanowires confirms the expected superior quality of the phase pure crystal structure, which is highly relevant for the reproducible definition of confined quantum states. The results obtained here are also important for future superconductor/semiconductor nanowire hybrid structures, e.g. for topological \cite{Oreg2010,Lutchyn2010,Prada2020} or Andreev-level qubits \cite{Zazunov2003, Tosi2019,Metzger2021,Zellekens2022}. An important aspect in this context is the coupling of the superconducting shell with the electron gas at the interface, which is favorably achieved by confining the electron system in the shell. In fact, when coupled to a superconducting contact, enhanced magnetoconductance oscillations with a period of $h/2e$ are observed, which are attributed to phase-coherent resonant Andreev reflections \cite{Guel2014a}. More recently, $h/2e$ oscillations have been found in the switching current of Josephson junctions based on GaAs/InAs core/shell nanowires \cite{Zellekens2024}. In light of these recent experiments, we believe that phase-pure semiconductor nanowires present a very interesting new research direction towards optimization of such hybrid devices.

\section*{Acknowledgements}

We thank Herbert Kertz for technical assistance. Dr. Florian Lentz and Dr. Stefan Trellenkamp are gratefully acknowledged for their help with the required e-beam lithography. The sample fabrication has been performed in the Helmholtz Nano Facility at Forschungszentrum J\"ulich \cite{Albrecht2017}. This work was partly funded by Deutsche Forschungsgemeinschaft (DFG, German Research Foundation) under Germany’s Excellence Strategy—Cluster of Excellence Matter and Light for Quantum Computing (ML4Q) EXC 2004/1—390534769. K.M.\ acknowledges the financial support by the Bavarian Ministry of Economic Affairs, Regional Development and Energy within Bavaria’s High-Tech Agenda Project "Bausteine für das Quantencomputing auf Basis topologischer Materialien mit experimentellen und theoretischen Ansätzen" (Grant No.\ 07 02/686 58/1/21 1/22 2/23) and by the German Federal Ministry of Education and Research (BMBF) via the Quantum Future project ‘MajoranaChips’ (Grant No.\ 13N15264) within the funding program Photonic Research Germany. A.M.S. and R.J. acknowledge the support of the EPSRC Grant EP/W002418/1. 

\putbib[bu1.bbl]  
\end{bibunit}



\clearpage
\widetext

\titleformat{\section}[hang]{\bfseries}{\MakeUppercase{Supplemental Note} \thesection:\ }{0pt}{\MakeUppercase}
\begin{bibunit}[]
\setcounter{section}{0}
\setcounter{equation}{0}
\setcounter{figure}{0}
\setcounter{table}{0}
\setcounter{page}{1}
\renewcommand{\thesection}{\arabic{section}}
\renewcommand{\thesubsection}{\Alph{subsection}}
\renewcommand{\theequation}{S\arabic{equation}}
\renewcommand{\thefigure}{S\arabic{figure}}
\renewcommand{\figurename}{Supplemental Figure}
\renewcommand{\tablename}{Supplemental Table}
\renewcommand{\bibnumfmt}[1]{[S#1]}
\renewcommand{\citenumfont}[1]{S#1}

\begin{center}
\textbf{\large Supporting Information: Aharonov--Bohm and Altshuler--Aronov--Spivak oscillations in the quasi-ballistic regime in phase-pure GaAs/InAs core/shell nanowires}
\end{center}

{
  \hypersetup{linkcolor=black}
  \tableofcontents
}


\section{Experimental Details}

\noindent The MBE growth of hexagonal-, or cubic-phase-pure GaAs core nanowires via self-catalysed vapour-liquid-solid technique on pre-structured substrates requires precise and dynamic control over the catalyst droplet on top of the growing nanowires. The evolution of the droplets contact angle over time determines the boundaries of axial phase changes along the nanowires. Following the recently developed approach of Jansen \textit{et al.} \cite{Jansen2020}, a contact angle of the Ga catalyst droplet around $110^\circ$ is appropriate to favour the wurtzite (WZ) phase or alternatively, above $125^\circ$, to aim for the zincblende (ZB) phase of the GaAs core. In Ref.~\cite{Jansen2020} we achieved the stabilization of the contact angle around $110^\circ$ by a dynamical reduction of the Ga flux by about $40$ \% during a growth time of 90 min. However, this WZ phase stabilization regime is extremely sensitive to the properties and precision of pre-structuring the substrates. For the as-grown nanowires we used Si(111) substrates covered with about 20\,nm of thermally deposited Si$\mathrm{O}_{2}$ hole arrays with varying diameters of 40, 60, and 80\,nm and pitches with 0.5, 1, 2, and 4\,$\upmu$m pinhole distances. These were predefined by electron beam lithography and subsequent dry and wet etching of the $\mathrm{SiO}_{2}$ down to a remaining thickness below 1\,nm. Note that precision of hole aspect ratio and etching depth is mainly limited by the etching processes used and can induce statistical deviations from the above-mentioned ideal stabilization conditions of the Ga catalyst droplet. After preparation, the pre-structured substrates are loaded into the load-lock of our MBE system and baked-out for about 45 min at $700^\circ$C. Then, the samples are transferred to the III/V-MBE chamber, heated to a substrate temperature of about $610^\circ$C and Ga is deposited for 10 min at a Ga beam equivalent pressure (BEP) of about $1.5\times 10^{-7}$\,mbar. During this step the Ga-droplets in the holes are formed, which under optimum conditions of the above-mentioned pre-structuring yield a starting contact angle around $90-95^\circ$. Following the pre-deposition of Ga, the growth of the GaAs core is initiated by applying an As flux with a BEP of $5\times 10^{-6}$ mbar for 90 min at the same substrate temperature, while the Ga flux is dynamically decreased to 60\,\% of the starting value. With the present growth conditions, a contact angle above $125^\circ$ is achieved, resulting in a zincblende crystal phase. The GaAs core growth is finalized with a 20\,min long droplet consumption step under reduced As BEP of $1\times 10^{-6}$ while the substrate temperature is kept at about $610^\circ$C. The InAs shell growth of the first growth run (samples A and B) is performed in total for 25\,min divided into 5 cycles, each one with 5\,min of growth alternated by 5\,min of growth break under As-stabilization. The corresponding substrate temperature is $450^\circ$C. In and As were supplied with BEPs of $1.95\times 10^{-7}$ mbar and $5\times 10^{-6}$ mbar, respectively. In case of the second growth run (sample C) the shell was grown for 40 \,min with according growth breaks, i.e. 8 cycles instead of 5. 


\section{Transmission Electron Microscopy Studies}

\begin{figure*}[!h]
	\centering
\includegraphics[width=1.0\linewidth]{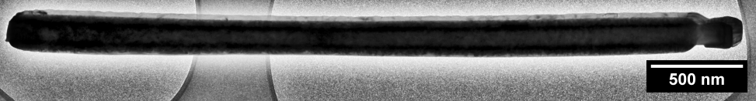}
	\caption{\justifying Bright-field TEM image of a nanowire of the first growth run with a diameter in the middle section of the wire of 210\,nm, corresponding to an example of the same growth batch as in the case of samples A and B.}
	\label{ig_Supp:Supp_Fig_S1}
\end{figure*}

\noindent In order to analyze the crystal structures at different locations along the nanowire by transmission electron microscopy (TEM), they are transferred onto a lacey carbon grid. The nanowires were examined by a JEOL 2100 TEM operated at 200\,keV. Supplemental Figure~\ref{ig_Supp:Supp_Fig_S1} shows a bright-field transmission electron micrograph (BF TEM) of a typical nanowire of the first growth run (representative samples A and B) 
with a length of 3.8\,$\upmu$m and diameter of 210\,nm in the middle part. 

\begin{figure}[!htbp]
    \centering 
    \includegraphics[width=1.0\textwidth]{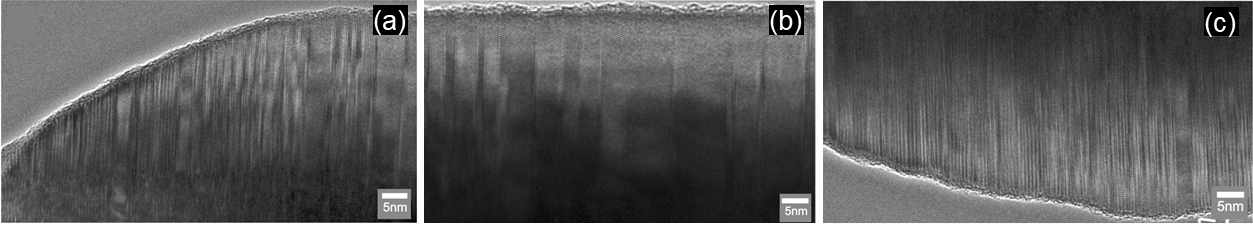}  
    \caption{\justifying (a) High-resolution TEM image of the top part of the nanowire shown in Supporting Figure~\ref{ig_Supp:Supp_Fig_S1}, implying the presence of wurtzite and zincblende segments. (b) The middle part of the nanowire shows longer zincblende segments with twins. (c) Bottom part indicating the presence of stacking faults and different crystal phases.}
	\label{ig_Supp:Supp_Fig_S2}
 \end{figure}

Supplemental Fig.~\ref{ig_Supp:Supp_Fig_S2} (a) shows a high-resolution TEM (HR TEM) image of the top part of the nanowire, confirming the presence of stacking faults and different crystal phases. Supplemental Fig.~\ref{ig_Supp:Supp_Fig_S2} (b) depicts the HR TEM image of the middle part of the nanowire showing only zincblende segments with twins. Finally, in Supplemental Fig.~\ref{ig_Supp:Supp_Fig_S2} (c), the bottom part is shown revealing once again stacking faults and different crystal phases.

\begin{figure*}[!h]
	\centering
\includegraphics[width=1.0\linewidth]{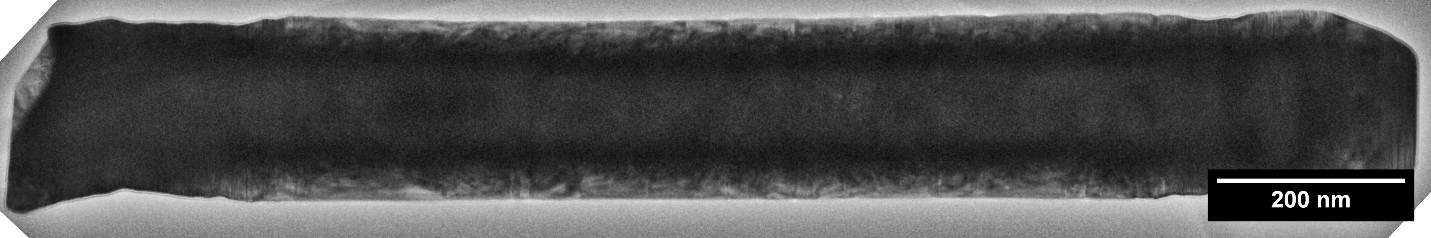}
	\caption{\justifying Bright-field TEM image of a nanowire of the second growth run with an average diameter of 188\,nm, corresponding to an example of the same growth batch as in the case of sample C.}
	\label{ig_Supp:Supp_Fig_S3}
\end{figure*}

Supplemental Fig.\,\ref{ig_Supp:Supp_Fig_S3} shows a BF TEM micrograph of a typical nanowire of the second growth run (representative sample C) with a length of 1.5\,$\upmu$m, and average diameter of 188\,nm. 

\begin{figure*}[!h]
	\centering
\includegraphics[width=1.0\linewidth]{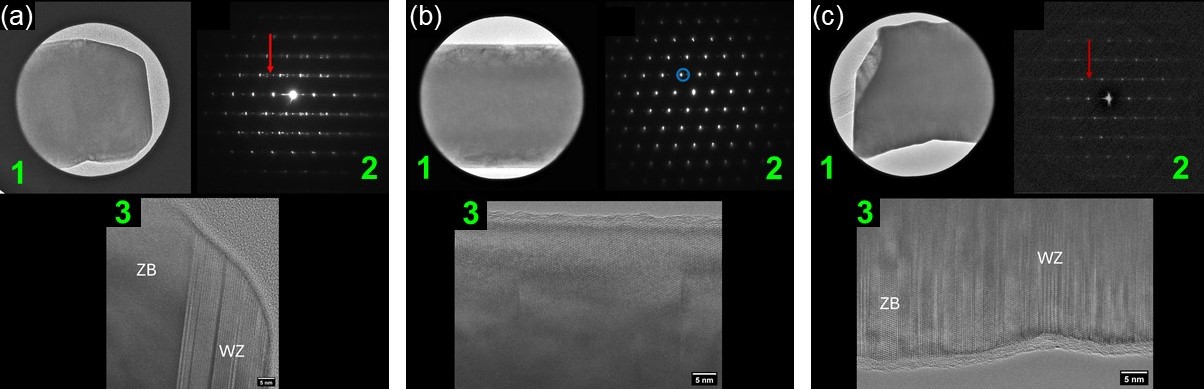}
	\caption{\justifying TEM images of the (a) top, (b) middle, and (c) bottom parts of a nanowire of the second growth run, each containing (1) the bright-field TEM image, (2) the corresponding SAED pattern, and (3)  the high-resolution TEM image.}
	\label{fig_Supp:Supp_Fig_S4}
\end{figure*}

Supplemental Fig.~\ref{fig_Supp:Supp_Fig_S4} (a) shows the selected-area image (1) of the nanowire top part and (2) is the corresponding selective-area electron diffraction (SAED) pattern. The streaky pattern (indicated by a red arrow in (2)), suggests the presence of stacking faults and different crystal phases. Subfigure (3) is the HR TEM image of the top part of the nanowire confirming the presence of WZ and ZB segments. A number of WZ segments were observed with a length of 18 and 11\,nm.

In Supplemental Fig.~\ref{fig_Supp:Supp_Fig_S4} (b) (1), the selected-area image of the middle part of the nanowire is shown with the corresponding SAED pattern given in (2). The streaky pattern is not observed in the SAED pattern. A pair of spots are observed (indicated by blue circle), suggesting the presence of twins and the absence of streaks indicate longer ZB segments. In Subfigure (3), showing the corresponding HR TEM image, ZB segments with twins can be identified.  

Supplemental Fig.~\ref{fig_Supp:Supp_Fig_S4} (c) (1) is the selected-area image of the nanowire bottom, and (2) is the corresponding diffraction pattern. The streaky pattern (indicated by a red arrow) indicates the presence of stacking faults and different crystal phases. In the HR TEM image given Subfigure (3), stacking faults along with ZB and WZ segments are revealed, i.e. supporting the observed diffraction pattern. A WZ segment was found at the bottom of the nanowire with a length of 5\,nm. 

\section{Additional Magnetotransport Measurements}

\subsection{Sample A}

The temperature-dependent normalized magnetoconductance $G/G_0$ measurements on sample A are presented in Supplemental Fig.~\ref{fig_Supp:Figure_S5} (a), with $G_0=e^2/h$. The conductance $G$ was measured as a function of axial magnetic field in the range of $-4$ to 4\,T for temperatures between 1.4 and 30\,K. Regular oscillations in the conductance signal with an Aharonov--Bohm (AB) period of 0.4\,T are observed. From this measurement it can be concluded that periodic oscillations persist up to about 30\,K. This agrees well with the fast Fourier transform (FFT) data analysis of the measurement (see Supplemental Fig.~\ref{fig_Supp:Figure_S5} (b)), where a $h/e$ peak is observed at a frequency of $2. 4\,\mathrm{T^{-1}}$, and a second peak corresponding to $h/2e$ periodic oscillations is found at 4.8$\,\mathrm{T^{-1}}$, which fits well with the observed values in the gate-dependent measurements presented in the main text.

\begin{figure*}[htb]
    \centering
    \includegraphics[width=1.0\linewidth]{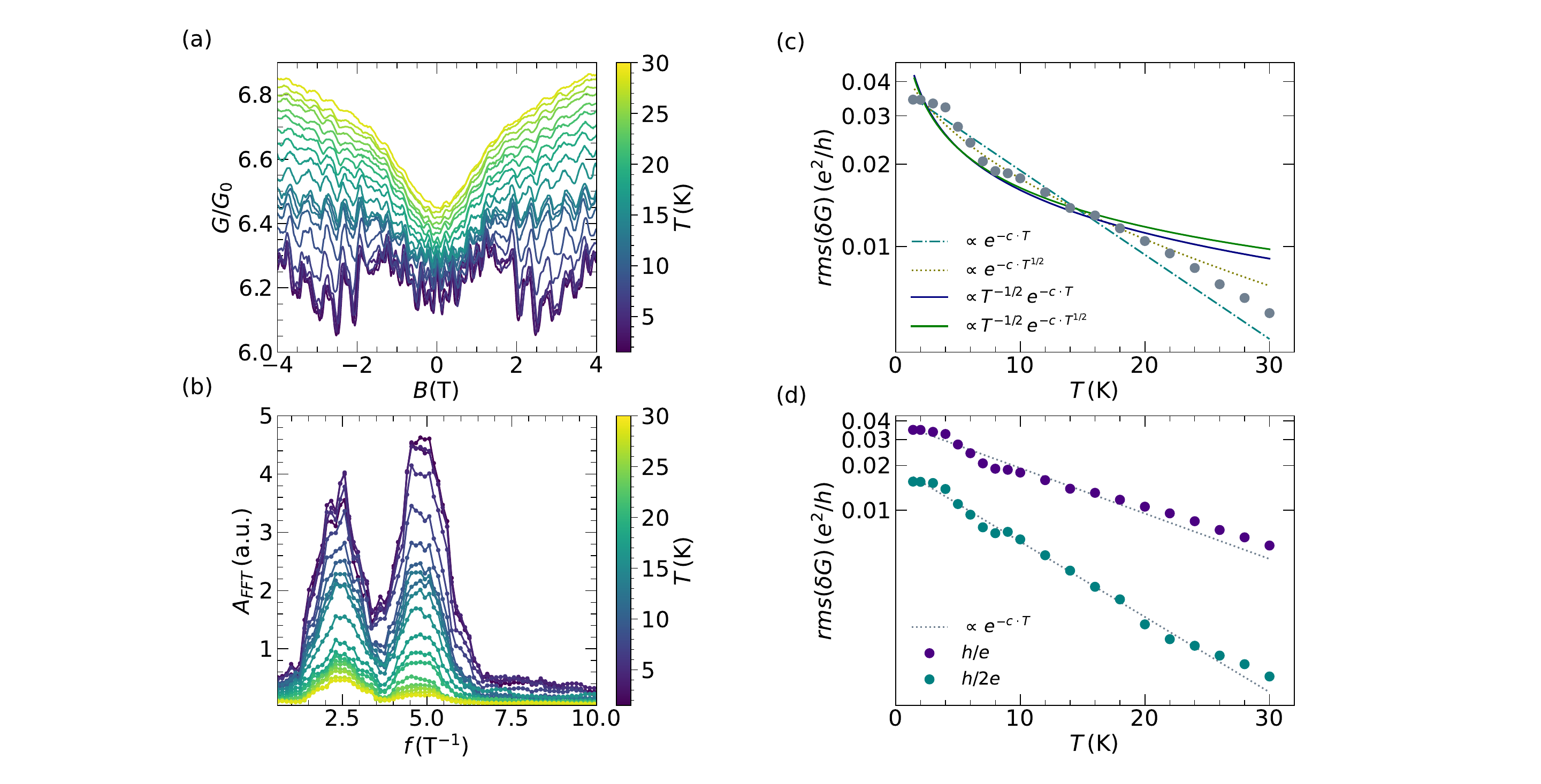}
	\caption{\justifying (a) Normalized conductance $G/G_{0}$ of sample A as the function of an axial magnetic field $B$ with varying temperature. It resolves periodic oscillations persistent up to around $30\,$K. (b) Corresponding fast Fourier transform data analysis, with pronounced frequency peaks belonging to $h/e$- and $h/2e$-periodic oscillations. (c) Extracted oscillation amplitude as a function of temperature, shown for different fits describing the diffusive and quasi-ballistic transport regime, with or without thermal broadening. (d) Quasi-ballistic fit without thermal broadening for $h/e$-, and $h/2e$-periodic oscillation amplitudes.}
	\label{fig_Supp:Figure_S5}
\end{figure*}

The temperature-dependent magnetoconductance measurement on sample A allows a detailed analysis of the transport regime by estimating the temperature dependence of the AB oscillation amplitude $\mathrm{rms}(\delta G)$ \cite{Webb86,Seelig2001}. Supplemental Fig.~\ref{fig_Supp:Figure_S5} (c), shows different fits corresponding to cases of transport, i.e. given with a prefactor $T^{-1/2}$ to account for when the thermal energy $k_{B}T$ is larger  than the Thouless energy, or without thermal broadening, as well as diffusive $\mathrm{exp}(-aT^{1/2})$, and quasi-ballistic $\mathrm{exp}(-aT)$ transport regime. It can be concluded that the quasi-ballistic fit without thermal broadening follows our data best.

Additionally, Supplemental Fig.~\ref{fig_Supp:Figure_S5} (d) demonstrates the quasi-ballistic fit for both $h/e$- and $h/2e$-periodic oscillations by extracting the $\mathrm{rms}(\delta G)$, using the inverse FFT method with pre-defined frequency windows for $h/e$-, and $h/2e$-periodic oscillation peaks. From the data presented, it can be concluded that such a fit can well describe both, $h/e$, and $h/2e$ conductance contributions in our data. In case of $h/e$ conductance contribution temperature dependency, a fit parameter of value $c=0.07\,\mathrm{K}^{-1}$, was extracted, and the phase-coherence length was estimated to be 3.3\,$\upmu$m.

\subsection{Sample B}

As for sample A, the analysis of the temperature-dependent magnetoconductance measurement results of sample B is discussed in the following text. The normalized conductance $G/G_{0}$ is shown as a function of axially applied magnetic field in the Supplemental Fig.~\ref{fig_Supp:Figure_S6} (a). Pronounced AB-type oscillations with a periodicity of 0.3\,T are observed, which persist up to about 30\,K. A beating pattern is observed in the magnetoconductance up to a temperature of about 12\,K. We attribute this to the presence of Zeeman splitting \cite{Rosdahl2014}. The oscillation period agrees well with the position of the frequency peak at 3.5$\,\mathrm{T^{-1}}$ in the FFT analysis, shown in Supplemental Fig.~\ref{fig_Supp:Figure_S6} (b). This value corresponds to the mean radius of the coherent trajectory of 67\,nm. A relatively small second peak is observed at about 6.6$\,\mathrm{T^{-1}}$, which vanishes around $5\,$K. Conducting the same transport regime analysis, by evaluating the AB oscillation amplitude $\mathrm{rms}(\delta G)$ temperature dependency, as shown in Supplemental Fig.~\ref{fig_Supp:Figure_S6} (c), we could conclude that the quasi-ballistic fit describes our data very well. From this, fitting parameter $c=0.1\,\mathrm{K}^{-1}$ was extracted, and the phase-coherence was estimated as $2.7\,\mathrm{\upmu m}$. 

\begin{figure*}[htb]
    \centering
    \includegraphics[width=1.0\linewidth]{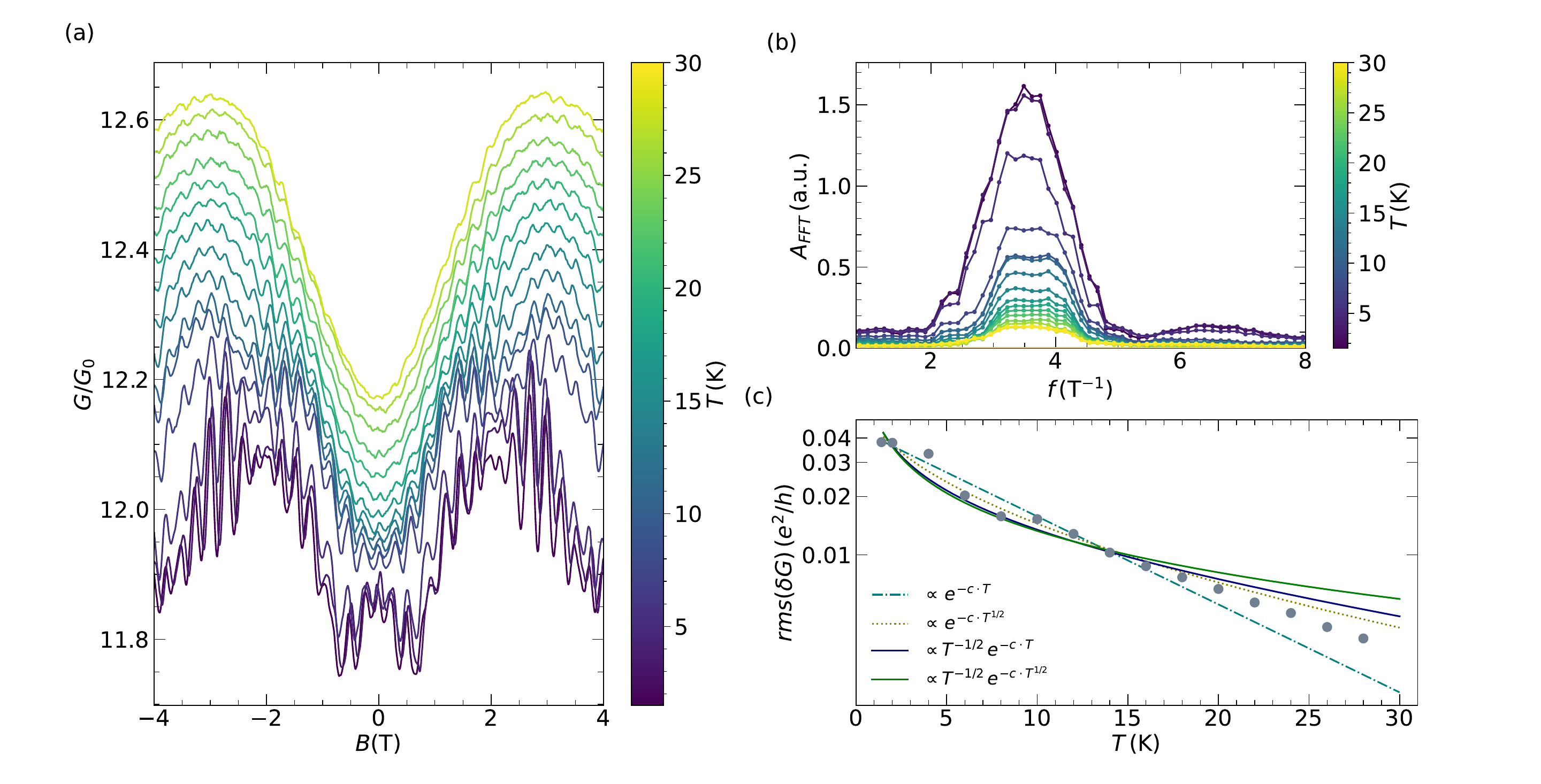}
	\caption{\justifying (a) Normalized conductance $G/G_{0}$ of sample B plotted as a function of magnetic field and recorded at temperatures in the range between 1.5 and 30\,K. It displays periodic oscillations for the entire temperature range. The conductance values are successively offset by 0.02$\,e^{2}/h$ for clarity. (b) Fourier transform analysis for the data shown in (a), using the oscillation signal $\delta G=G-\overline{G}$, in order to subtract varying background contribution. It shows a $h/e$-periodic oscillations peak, together with a weak and quickly vanishing $h/2e$-periodic oscillations peak. (c) Transport regime analysis shown as in case of sample A. It shows that the quasi-ballistic fit is the optimal way to describe the transport regime in sample B.}
	\label{fig_Supp:Figure_S6}
\end{figure*}

In Supplemental Fig.~\ref{fig_Supp:Figure_S7} (a) the magnetoconductance is shown for different values of the applied gate voltage $V_\mathrm{g}$ ranging from $-3$ to 3\,V. The axial magnetic field was applied from $-2$ to 2\,T. Regular oscillations are seen with varying conductance minima and maxima at $B=0$. The corresponding FFT analysis of this data is shown in Supplemental Fig.~\ref{fig_Supp:Figure_S7} (b). In this measurement, no signs of $h/2e$ conductance contributation is found. The gate-dependent measurement shown here was performed after the temperature-dependent one, so it can be argued that due to multiple thermal cycles, the potential landscape in the nanowire has changed significantly, leading to absence of such contribution. To further illustrate the evolution of the $h/e$ periodic oscillations across the gate voltage range, the filtered out $h/e$ conductance contribution is given by the colormap in Supplemental Fig.~\ref{fig_Supp:Figure_S7} (c). Phase jumps between 0 and $\pi$ are clearly observed at $B=0$. 

\begin{figure*}[htb]
    \centering
    \includegraphics[width=0.85\linewidth]{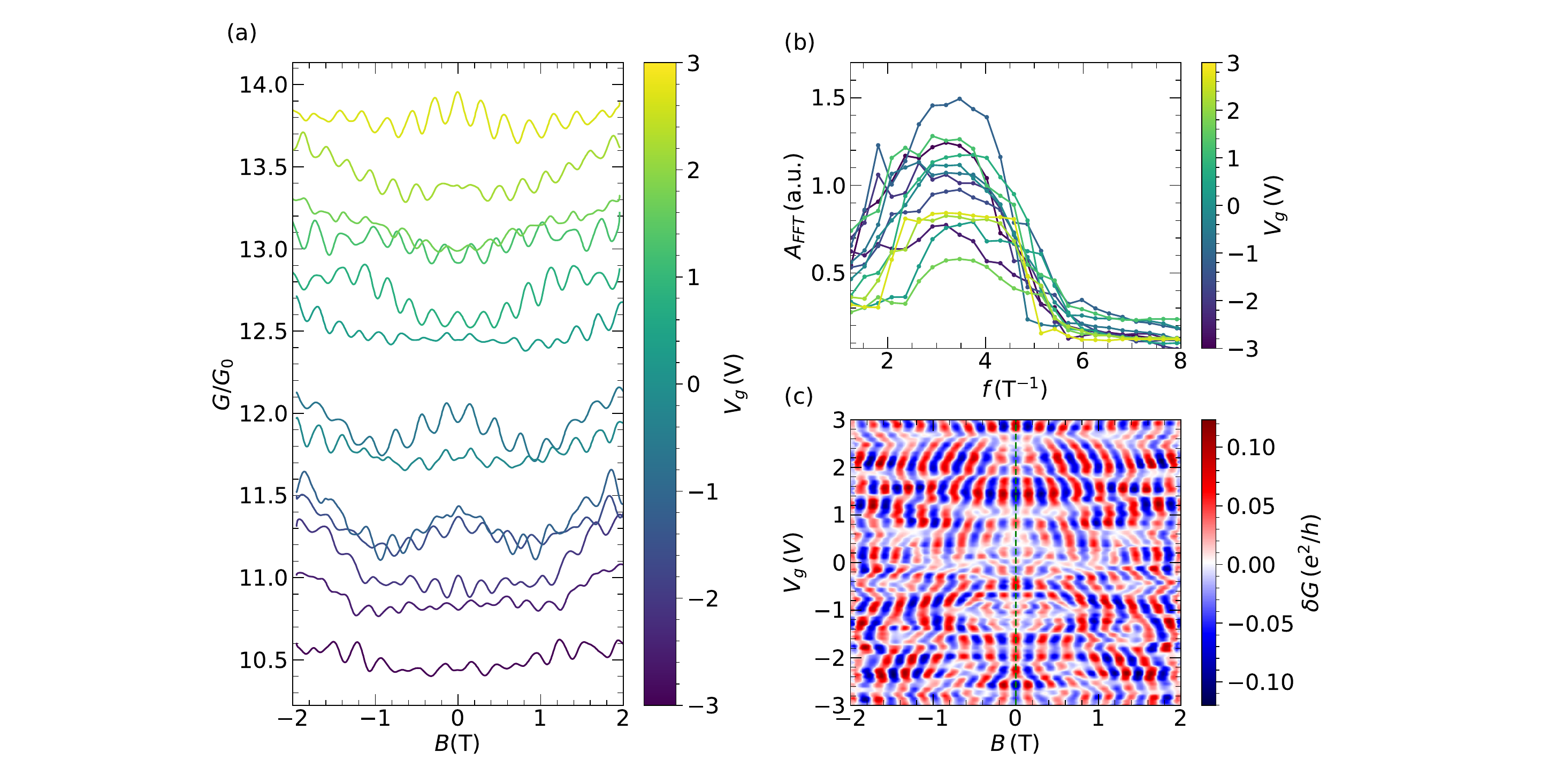}
	\caption{\justifying (a) Normalized magnetoconductance $G/G_{0}$ for several values of applied gate voltage. The AB oscillations persist in the entire gate voltage range. The conductance values are successively offset by 0.01$\,e^{2}/h$ for clarity. (b) FFT analysis of the subtracted data in (a), corresponding to $\delta G$. It shows only a $h/e$-periodic contribution, without the occurrence of the second peak as for other samples. (c) Filtered $h/e$-periodic oscillations as a function of gate voltage and magnetic field, featuring oscillation amplitude jumps along the $B=0$ line as a consequence of the gate-dependent phase.}
	\label{fig_Supp:Figure_S7}
\end{figure*}

\subsection{Sample C}

\begin{figure*}[htb]
    \centering
    \includegraphics[width=0.95\linewidth]{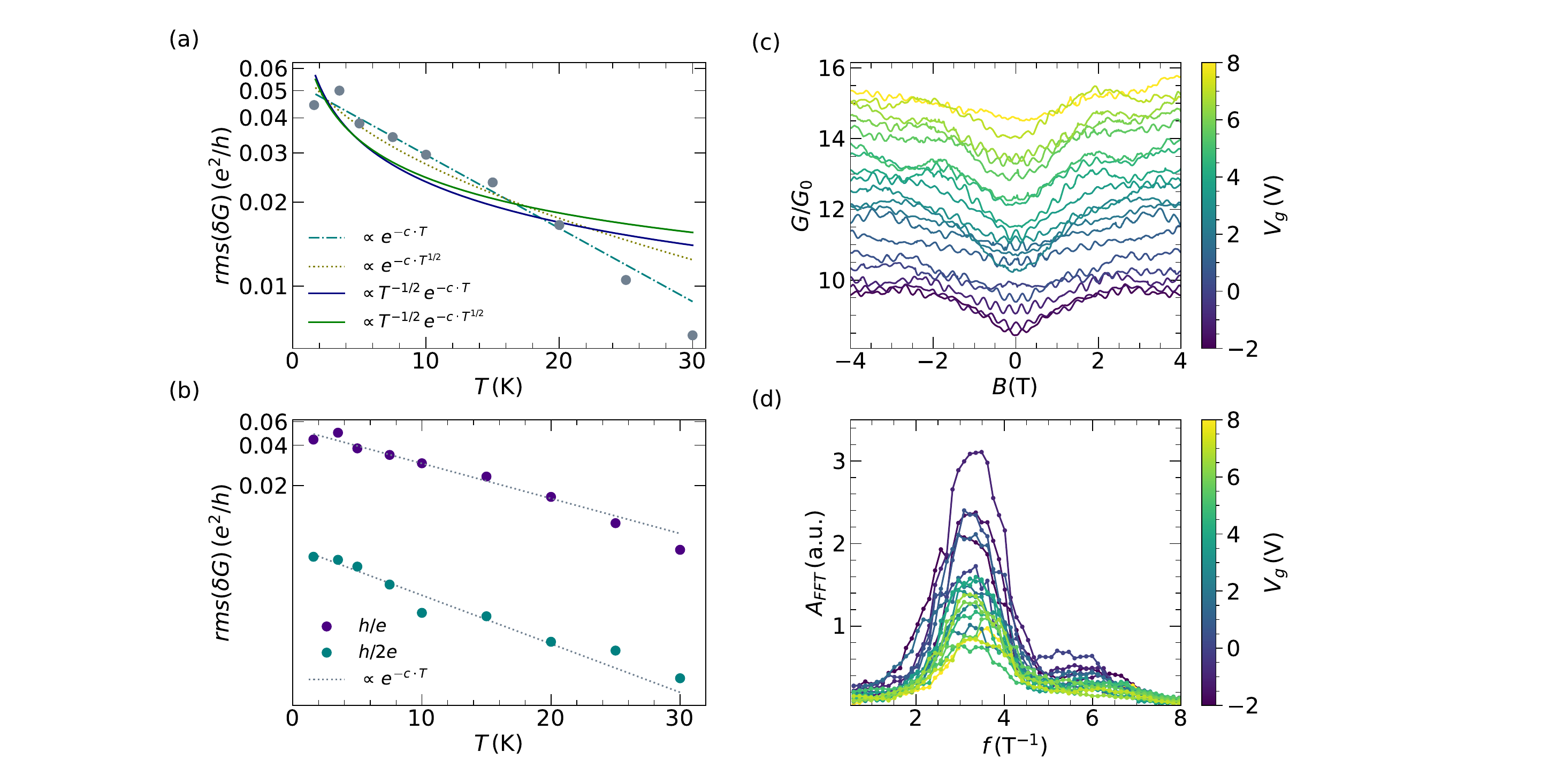}
	\caption{\justifying (a) Root-mean-square $\mathrm{rms}(\delta G)$ of the AB oscillation amplitude as a function of temperature, shown together with fits describing different transport regimes (see sections A and B for more information). (b) Quasi-ballistic fit for the $h/e$-, and $h/2e$-periodic oscillations. (c) Normalized magnetoconductance of Sample C for various gate voltages $V_{g}$. It features persistent oscillations for the entire gate voltage range. (d) Corresponding fast Fourier transform of the magnetoconductance shows two-peak structure, with $h/2e$-periodic contribution vanishing as the applied gate voltage increases.}
	\label{fig_Supp:Figure_S8}
\end{figure*}

In the main text, a detailed analysis of temperature-dependent magnetoconductance measurements for sample C is given. As an additional support to the transport regime evaluation, Supplemental Fig.~\ref{fig_Supp:Figure_S8} (a) features four different transport regime fits, as described in the previous sections. This graph demonstrates the quasi-ballistic fit as a good way to describe the transport regime in our sample. From the quasi-ballistic fit, using the extracted fitting parameter $c=0.06$, the phase coherence length was determined to be $4.6\,\mathrm{\upmu m}$. In addition, Supplemental Fig.~\ref{fig_Supp:Figure_S8} (b) shows that such a fit describes both $h/e$ and $h/2e$ conductance contributions well.

Furthermore, Supplemental Fig.~\ref{fig_Supp:Figure_S8} (c) shows the gate-dependent magnetoconductance for applied gate voltage in the range from $-2$ to $8$\,V in steps of 0.25\,V at a temperature of 1.5\,K. Once again, persistent AB oscillations are observed for the entire gate voltage range. Similar to the case of temperature-dependent measurement, a $h/e$-periodic contribution peak in FFT analysis (Supplemental Fig.~\ref{fig_Supp:Figure_S8} (d)) is observed at $3.3\,\mathrm{T}^{-1}$, whereas the second peak, corresponding to the $h/2e$ contribution, is found at $5.8\,\mathrm{T}^{-1}$, and vanishes around $3\,$V.

\afterpage{\clearpage}
\clearpage

\section{Modelling using Kubo Formalism\label{section:Kubo}}

Using the Kubo formalism, the magnetoconductance was calculated for a nanowire with geometrical dimensions similar to those of sample A. The nanowire was described as a tubular shell with infinite length, and with an external radius $R=r_{\mathrm {tot}}$. The InAs material was represented by the effective mass $0.023\,m_0$ and by the $g$ factor of $-14.9$. The transverse section was discretized on a polar lattice \cite{Sitek2015,Manolescu2017}.  The eigenstates of the Hamiltonian were obtained by numerical diagonalization and were labeled as $|nks\rangle$, where $k$ is the wave vector of the plane waves in the longitudinal direction $z$, $n=0,1,2,...$ labels the transverse modes, and $s=\pm 1$ is the spin index.  For this nanowire model the conductivity was calculated with the linear-response Kubo formula,
\begin{equation}
\sigma_{zz} = \   \frac{e^2}{A} \frac{\hbar}{\pi^2}\!
\int \! dE \left[-\frac{\partial \cal F} {\partial E}\right]\ \int \! dk \
\!\! \sum_{\substack{n_1 s_1 \\ n_2 s_2} } \!
|\langle n_1 k s_1 |v_z|n_2 k s_2\rangle |^2 \
\mathrm {Im} \ G_{n_1 k s_1}(E) \ \mathrm {Im} \ G_{n_2 k s_2}(E) \ ,
\label{kuboc}
\end{equation}
where $A$ is the transverse area of the shell conductor, $v_z$ the velocity operator in the $z$-direction, $\cal F$ the Fermi function, and $G_{nks}(E)=\left[(E-E_{nks}-\Sigma(E)\right]^{-1}$ the Green's function \cite{Urbaneja2018}.  The nanowire includes disorder implemented as uniformly distributed point-like impurities, which generate the energy dependent self-energy $\Sigma(E)$. In the present simple approximation $\Sigma(E)$ is considered a complex number, which is obtained by an averaging procedure over all possible impurity configurations, in the self-consistent Born approximation \cite{Doniach1998}. That procedure leads to solving the equation 
\begin{equation}
\Sigma(E) = \gamma^2 \sum_{n s} \int \frac{d(kR)}{E-E_{nks}-\Sigma(E)} \ ,
\label{scba}
\end{equation}
with $\gamma$ being an energy parameter associated with the impurities.

To compare with the experimental data we converted the conductivity to conductance, by multiplying Eq.\ (\ref{kuboc}) with the factor $A/L_c$, where $L_c$ is the length of the nanowire. In the main text the normalized magnetoconductance is presented as a function of the chemical potential (see e.g. Fig. 3 (a) in the main text), for disorder potential $0.18\,$meV and temperature $1.5\,$K. The change in chemical potential is directly related to the change in gate voltage in the experiment. With increasing gate voltage and thus increasing carrier concentration, the chemical potential increases accordingly. The obtained Kubo formalism calculations capture the basic features of the measurements, where the magnetoconductance increases with increasing chemical potential and oscillates with the period of a magnetic flux quantum $\Phi_0$. The phase of the oscillations changes with the chemical potential, which was also observed in the experiment, see e.g. Fig.~2 (c) in the main text. 

\begin{figure*}[htb]
    \centering
    \includegraphics[width=0.95\linewidth]{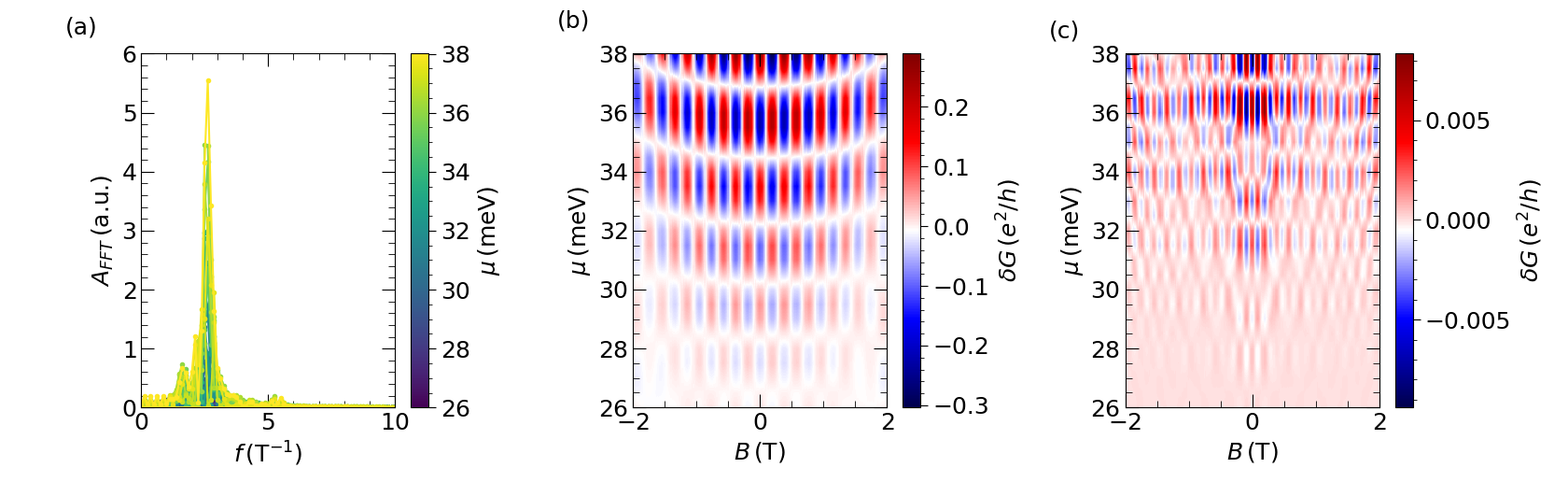}
	\caption{\justifying (a) FFT spectrum for the data presented in the main text by Fig.~3 (a), using the calculated $\delta G$ values. Peaks corresponding to the $h/e$ conductance contribution are found at $2.5\,\mathrm{T}^{-1}$, and for $h/2e$ at $5.3\,\mathrm{T}^{-1}$. (b) Color map of the isolated $h/e$ conductance contribution. It reveals phase switching over the $B=0$ line. (c) Color map of the isolated $h/2e$ conductance contribution. For this case, the experimentally observed phase rigidity at $B=0$ line is not observed due to nature of the Kubo calculation routine and its limitations to accurately describe our system.}
	\label{fig_Supp:Figure_S9}
\end{figure*}

Additional to the normalized magnetoconductance plot shown in main text, Supplemental Fig.~\ref{fig_Supp:Figure_S9} (a) displays the FFT analysis of the presented data with subtracted background ($\delta G=G-\overline{G}$) for a chemical potential ranging from 26 to 38\,meV. These calculations feature similar $h/e$ and $h/2e$ peak positions on the frequency scale, as compared to the experimental values of sample A. 

By individually analysing these frequency windows, a chemical potential-dependent phase analysis could be made, following the same method as for our experimental data. In Supplemental Fig.~\ref{fig_Supp:Figure_S9} (b), the $h/e$ contribution is presented, showing the expected phase switching over the entire chemical potential range along the $B=0$ line. However, the same analysis of the isolated $h/2e$ contribution presented in Supplemental Fig.~\ref{fig_Supp:Figure_S9} (c), does not reveal any phase-rigidity at the $B=0$ line. In fact, in the calculations using the Kubo formalism, higher order terms are not included, so we do not expect any phase rigid contributions arising from localization effects. We therefore attribute the observed second harmonic in the FFT to non-sinusoidal contributions of the Aharonov--Bohm type oscillations. 

\section{Tight Binding Simulations using KWANT \label{section:KWANT}}

A tight binding model representing the GaAs/InAs core/shell nanowire system is created with the package Kwant \cite{Groth2014}. This model is an adaptation of the model presented in Ref.~\cite{Bommer2019}, starting from the following continuum model Hamiltonian: 
\begin{eqnarray}
        H &=& \left[ \frac{\hbar^2 (k_x^2 + k_y^2 + k_z^2)}{2m_\mathrm{eff}}  - \mu + V(x, y, z) \right] \sigma_0  \nonumber \\
        && - \alpha \left\{ k_x \left[ (\sigma_z \sin(\theta_\mathrm{so}) + \sigma_y \cos(\theta_\mathrm{so})\right] 
        +  \sigma_x \left[ k_y \cos(\theta_\mathrm{so}) + k_z \sin(\theta_\mathrm{so}) \right] \right\} \nonumber \\
        && + \frac{1}{2}  g  \mu_\mathrm{B} (B_x  \sigma_x + B_y  \sigma_y + B_z  \sigma_z) \, ,
        \label{Eq:HamTB}
\end{eqnarray}
The Hamiltonian consists of three terms:
the first term describes the quadratic dispersion of a free electron gas with wave vector $\mathbf{k} = (k_x, k_y, k_z)$ and effective mass $m_\mathrm{eff}$, also including the chemical potential $\mu$ and potential $V(x,y,z)$ that describes all remaining potentials present within the model. Inside the core, this potential is taken to be 1~eV (neglecting disorder) while in the shell the potential represents the disorder, described by a Gaussian random field with disorder strength $S_\mathrm{dis} = 5\,\textnormal{meV}$ and correlation length $l_\mathrm{corr} = 50\,\textnormal{nm}$$: \langle V(\mathbf{r}) V(\mathbf{r}') \rangle = S_\mathrm{dis}^2 \exp[-(\mathbf{r} - \mathbf{r}')^2/(2l_\mathrm{corr}^2)]$. This correlation length is chosen rather comparable to the length of the wire (significantly exceeding the atomic scale), in order to represent a phase-pure nanowire in the quasi-ballistic regime with few scattering centers over the length of the transport section. An example of such a disorder profile is given in Supplemental Fig.~\ref{fig_Supp:Figure_S10}.

\begin{figure}[h]
    \centering
    \includegraphics[width=0.6\linewidth]{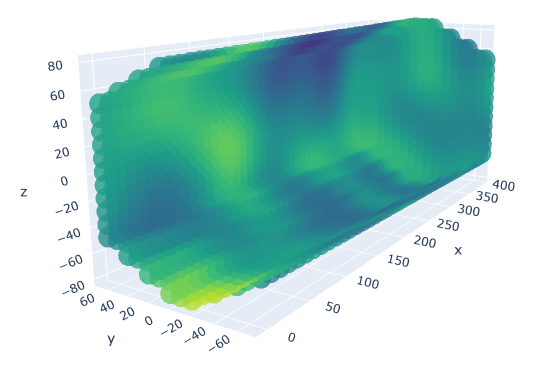}
    \caption{\justifying An example of a tight-binding model wire with hexagonal cross section and spatially correlated disorder potential illustrated by the green-blue color pattern. The coordinates of the sites are shown in units of nm.}
    \label{fig_Supp:Figure_S10}
\end{figure}

The second term of the Hamiltonian covers the spin-orbit coupling with respect to a specific surface, for which the surface normal is orthogonal to the $x$-axis. We take $\sigma_i$, with $i=x,y,z$, as the Pauli matrices acting on the spin (and $\sigma_0$ the identity matrix). In the spin-orbit term, $\alpha$ is the spin-orbit coupling strength, while the angle $\theta_\mathrm{so}$ depends on the location of the site in question. The angle is defined so that the spin-orbit coupling is given in the direction of the nearest surface facet, which is expected to be the dominant orientation of spin-orbit coupling in the nanowire under consideration.

The third term of Eq.~(\ref{Eq:HamTB}) is the Zeeman term, with $g$ the g-factor and $\mu_\mathrm{B}$ the Bohr magneton. The magnetic field under consideration is applied along the $x$-direction and given by $\mathbf{B}=(B, 0, 0)$, with corresponding vector potential $\mathbf{A}=(0, -z B, 0)$. This vector potential was implemented using the Peierls substitution method. 

The continuum model Hamiltonian is discretized using the finite difference method on a cubic lattice with a lattice constant that is sufficiently small to fit at least two layers of sites in the shell to represent the conducting modes enveloping the core. This discretization scheme is sufficient to capture the appearance of AB and AAS oscillations but is probably not sufficient to capture the loss of phase matching due to the finite thickness of the shell, as seen in experiment.

The model consists of a scattering region with two semi-infinite leads attached at opposite ends. The only difference between the Hamiltonian in the leads and the Hamiltonian in the scattering region is the added disorder in the scattering region as a function of $V(x,y,z)$.
The nanowire itself has a hexagonal shape for the core and shell, the approximate outer radii of which are 55\,nm and 85\,nm, respectively, comparable to the dimensions of sample A. The hexagonal shape can be seen in Supplemental Fig.~\ref{fig_Supp:Figure_S10}. The relevant simulation parameters are summarized in Supplemental Table~\ref{tab:nanowireparams}. 
\begin{table}[h]
    \centering
    \begin{tabular}{|c|c|}
    \hline
        Parameter & Value \\
        \hline
        $r_\mathrm{core}$ & $55$\,nm \\
        \hline
        $r_\mathrm{tot}$ & $85$\,nm \\
        \hline
        $l_\mathrm{corr}$ & $50$\,nm  \\
        \hline
        $a$ & 10\,nm \\
        \hline
        $m_\mathrm{eff}$ & $0.023\,m_\mathrm{e}$ \\
        \hline
        $g$ & $-14.9$ \\
        \hline
        $\mu$ & 14 - 45\,meV \\
        \hline
        $\alpha$ & 20\,meV$\cdot$nm\\
        \hline
        $V_\mathrm{core}$ & 1\,eV \\
        \hline
        $S_\mathrm{dis}$ & 5\,meV \\
        \hline
        $T$ & 1.7\,K \\
        \hline
    \end{tabular}
    \caption{\justifying The model parameters considered in the simulations are listed here: $r_\mathrm{core}$ and $r_\mathrm{tot}$ are the radius of the circle encircling the core and the radius of the circle encircling the shell, respectively; $l_\mathrm{corr}$ is the correlation length of the disorder profile in the nanowire shell; $a$ is the lattice constant of the tight binding model; $m_\mathrm{eff}$ is the effective mass and $m_\mathrm{e}$ the free electron mass; $g$ is the g-factor; $\mu$ the chemical potential; $\alpha$ the spin-orbit coupling strength; $V_\mathrm{core}$ the potential in the core of the nanowire; $S_\mathrm{dis}$ is the disorder strength of the disorder profile in the nanowire shell; the temperature at which the magnetoconductance is evaluated.}
    \label{tab:nanowireparams}
\end{table}

The magnetoconductance was obtained as a function of the magnetic field and chemical potential through the Landauer formula and scattering matrix calculations in Kwant:
\begin{equation} \label{Eq:Kwant-conductance}
    G(B, \mu) = \frac{e^2}{h} \int dE \left[ - \frac{\partial \mathcal{F}}{\partial E} \right] T(E, \mu, B),
\end{equation}
with Fermi function $\mathcal{F}$ and transmission probability $T(E, \mu, B)$ over the scattering region (summed over all channels). Note that energy and chemical potential can be interchanged, $T(E, \mu, B) = T(E + \mu, 0, B) = T(0, \mu + E, B)$, and furthermore $T(E, \mu, B) = T(E, \mu, -B)$, such that the finite-temperature conductance can be evaluated from transmission probabilities at zero energy, obtained over a range of chemical potentials and positive magnetic field strengths.

\begin{figure*}[htb]
    \centering
    \includegraphics[width=0.98\linewidth]{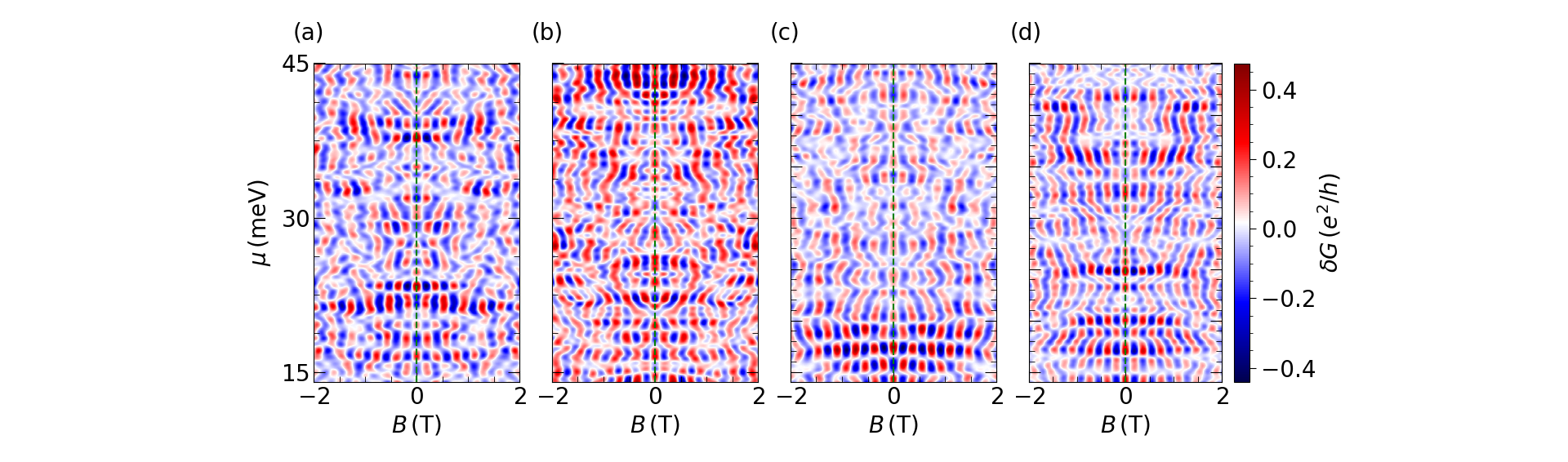}
	\caption{\justifying (a) to (d) Filtered $h/e$-periodic conductance contribution as a function of magnetic field $B$ and chemical potential $\mu$ of a 4\,$\upmu$m long wire with a disorder potential of 5\,meV at 1.7\,K. The conductance of each figure was determined by using a different set of disorder potentials, and shows phase switching along the $B=0$ line, regardless of the constructed potential landscape. The simulation presented in (a) is already discussed in the main text, and reproduced here for a comparison with remaining three sets presented in (b) to (d).}
	\label{fig_Supp:Figure_S11}
\end{figure*}

Supplemental Figs.~\ref{fig_Supp:Figure_S11} (a) to (d) show the filtered $h/e$-periodic conductance contributions as a function of magnetic field $B$ and chemical potential $\mu$. In total, four different sets of disorder potential configurations in InAs shell are considered, each with a disorder strength of 5\,meV, with the conductance evaluated at a temperature of 1.7\,K. The first of this set is presented and discussed in the main text, whereas the remaining three are briefly summed up in this section. 

For all disorder configurations we find that the phase at zero magnetic field switches between 0 and $\pi$. However, the chemical potential values at which this switching occurs are different for each configuration. The phase is balanced between 0 and $\pi$ with a small tendency towards a conductance peak, i.e. phase 0. Furthermore, the phase also switches randomly at finite magnetic fields.

\begin{figure*}[htb]
    \centering
    \includegraphics[width=0.98\linewidth]{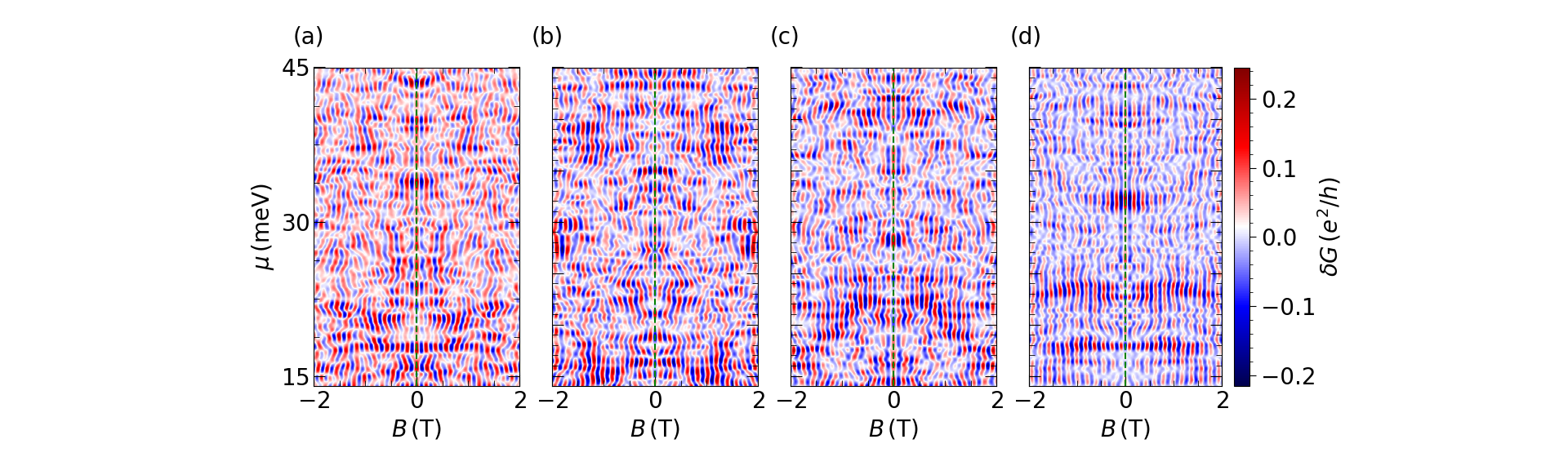}
	\caption{\justifying (a) to (d) Filtered $h/2e$-periodic contribution corresponding to the $h/e$-periodic contribution shown in Supplemental Fig.~\ref{fig_Supp:Figure_S11}. Here, one finds a dominating oscillation amplitude peak along the zero magnetic field line. The data set in (a) has already been presented in the main text and is shown here for comparison with the other three data sets.}
	\label{fig_Supp:Figure_S12}
\end{figure*}

Supplemental Figs.~\ref{fig_Supp:Figure_S12} (a) to (d) show, similarly to Supplemental Figs.~\ref{fig_Supp:Figure_S11}, the corresponding filtered $h/2e$-periodic conductance contributions. For all generated sets of disorder potential configurations, we predominantly find a peak at $B=0$, with only very short sections where the phase is switched. 
This finding is in agreement with our experimental observation of phase rigidity around zero magnetic field due to the occurrence of Altshuler--Aronov--Spivak oscillations in the presence of spin-orbit coupling. In the main text, the results for the different disorder configurations presented in Supplemental Figs.~\ref{fig_Supp:Figure_S12} and~\ref{fig_Supp:Figure_S11} are further discussed with respect to their average properties.

Finally, we also present the magnetoconductance results that are obtained for a nanowire segment that is only 400\,nm long, with all other simulation parameters kept fixed.
\begin{figure*}[htb]
    \centering
    \includegraphics[width=1.0\linewidth]{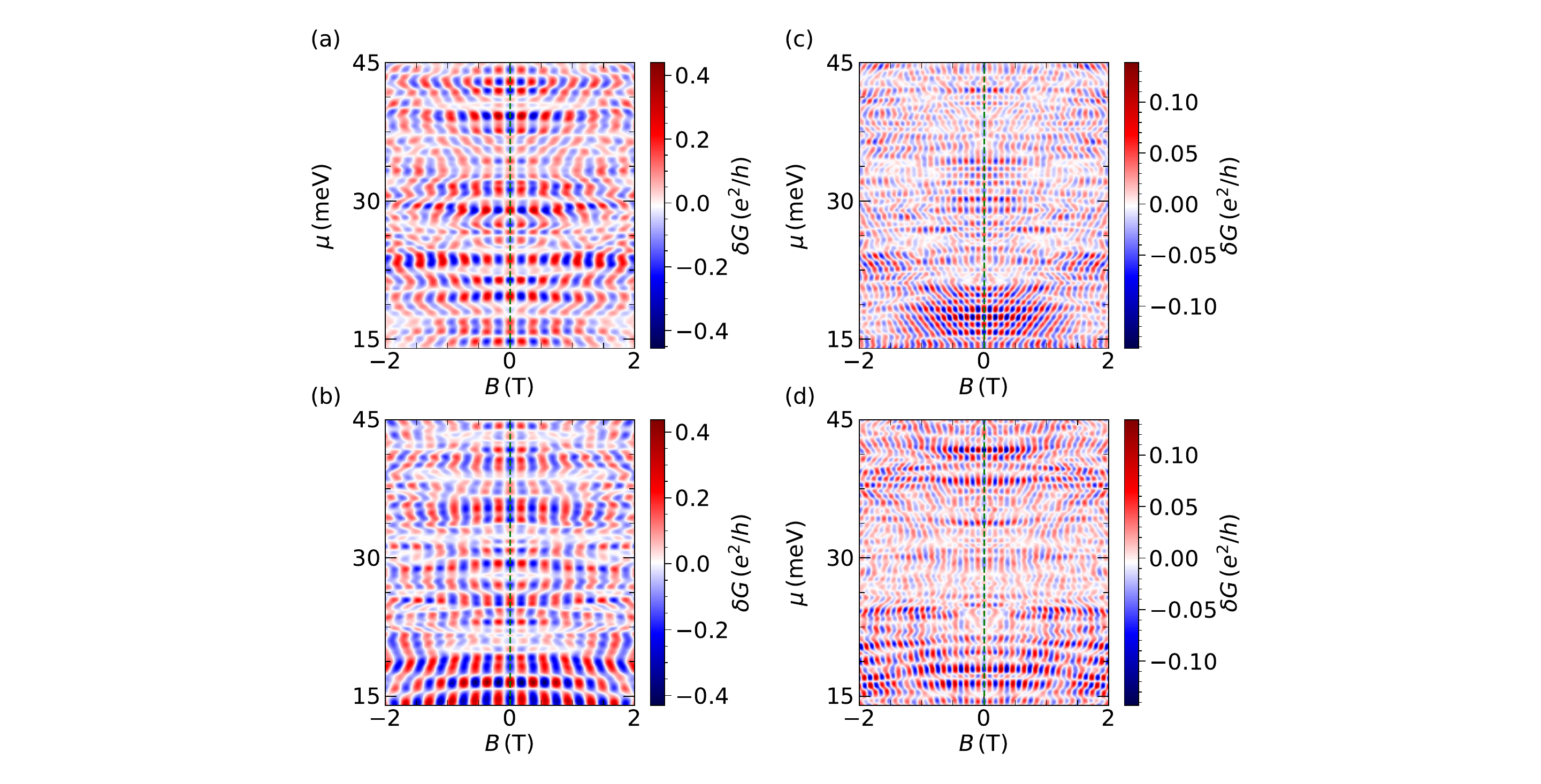}
	\caption{\justifying (a) and (b) Filtered $h/e$ conductance contribution for the case of a nanowire segment of 400\,nm length for two different disorder potential profiles. (c) and (d) Corresponding $h/2e$ contributions for the two sets shown in (a) and (b). They do not show a clear oscillation amplitude maximum at $B=0$, as in the case of calculations that include the entire nanowire length of 4\,$\upmu$m.}
	\label{fig_Supp:Figure_S13}
\end{figure*}
This distance corresponds to the inner contact separation, between which the voltage drop was recorded. In Supplemental Figs.~\ref{fig_Supp:Figure_S13} (a) and (b), examples of the isolated $h/e$ conductance contributions are presented for the case of two different sets of calculations comprising different potential landscape profiles. As expected, for both sets we observe phase switching along the $B=0$ line, associated with AB oscillations. However, for the corresponding filtered out $h/2e$ conductance contribution shown in Supplemental Figs.~\ref{fig_Supp:Figure_S13} (c) and (d), no prominent peak along the $B=0$ line is found. As discussed in the main text and in Supplementary Note~\ref{section:Kubo}, such a feature is more likely to represent a second harmonic contribution of the AB oscillations than of the AAS oscillations. 


\begin{thebibliography}{49}%
\makeatletter
\providecommand \@ifxundefined [1]{%
 \@ifx{#1\undefined}
}%
\providecommand \@ifnum [1]{%
 \ifnum #1\expandafter \@firstoftwo
 \else \expandafter \@secondoftwo
 \fi
}%
\providecommand \@ifx [1]{%
 \ifx #1\expandafter \@firstoftwo
 \else \expandafter \@secondoftwo
 \fi
}%
\providecommand \natexlab [1]{#1}%
\providecommand \enquote  [1]{``#1''}%
\providecommand \bibnamefont  [1]{#1}%
\providecommand \bibfnamefont [1]{#1}%
\providecommand \citenamefont [1]{#1}%
\providecommand \href@noop [0]{\@secondoftwo}%
\providecommand \href [0]{\begingroup \@sanitize@url \@href}%
\providecommand \@href[1]{\@@startlink{#1}\@@href}%
\providecommand \@@href[1]{\endgroup#1\@@endlink}%
\providecommand \@sanitize@url [0]{\catcode `\\12\catcode `\$12\catcode
  `\&12\catcode `\#12\catcode `\^12\catcode `\_12\catcode `\%12\relax}%
\providecommand \@@startlink[1]{}%
\providecommand \@@endlink[0]{}%
\providecommand \url  [0]{\begingroup\@sanitize@url \@url }%
\providecommand \@url [1]{\endgroup\@href {#1}{\urlprefix }}%
\providecommand \urlprefix  [0]{URL }%
\providecommand \Eprint [0]{\href }%
\providecommand \doibase [0]{https://doi.org/}%
\providecommand \selectlanguage [0]{\@gobble}%
\providecommand \bibinfo  [0]{\@secondoftwo}%
\providecommand \bibfield  [0]{\@secondoftwo}%
\providecommand \translation [1]{[#1]}%
\providecommand \BibitemOpen [0]{}%
\providecommand \bibitemStop [0]{}%
\providecommand \bibitemNoStop [0]{.\EOS\space}%
\providecommand \EOS [0]{\spacefactor3000\relax}%
\providecommand \BibitemShut  [1]{\csname bibitem#1\endcsname}%
\let\auto@bib@innerbib\@empty
\bibitem [{\citenamefont {Badawy}\ and\ \citenamefont
  {Bakkers}(2024)}]{Badawy2024}%
  \BibitemOpen
  \bibfield  {author} {\bibinfo {author} {\bibfnamefont {G.}~\bibnamefont
  {Badawy}}\ and\ \bibinfo {author} {\bibfnamefont {E.~P. A.~M.}\ \bibnamefont
  {Bakkers}},\ }\href {https://doi.org/10.1021/acs.chemrev.3c00656} {\bibfield
  {journal} {\bibinfo  {journal} {Chemical Reviews}\ }\textbf {\bibinfo
  {volume} {124}},\ \bibinfo {pages} {2419} (\bibinfo {year} {2024})},\
  \bibinfo {note} {pMID: 38394689}\BibitemShut {NoStop}%
\bibitem [{\citenamefont {de~Lange}\ \emph {et~al.}(2015)\citenamefont
  {de~Lange}, \citenamefont {van Heck}, \citenamefont {Bruno}, \citenamefont
  {van Woerkom}, \citenamefont {Geresdi}, \citenamefont {Plissard},
  \citenamefont {Bakkers}, \citenamefont {Akhmerov},\ and\ \citenamefont
  {DiCarlo}}]{deLange2015}%
  \BibitemOpen
  \bibfield  {author} {\bibinfo {author} {\bibfnamefont {G.}~\bibnamefont
  {de~Lange}}, \bibinfo {author} {\bibfnamefont {B.}~\bibnamefont {van Heck}},
  \bibinfo {author} {\bibfnamefont {A.}~\bibnamefont {Bruno}}, \bibinfo
  {author} {\bibfnamefont {D.~J.}\ \bibnamefont {van Woerkom}}, \bibinfo
  {author} {\bibfnamefont {A.}~\bibnamefont {Geresdi}}, \bibinfo {author}
  {\bibfnamefont {S.~R.}\ \bibnamefont {Plissard}}, \bibinfo {author}
  {\bibfnamefont {E.~P. A.~M.}\ \bibnamefont {Bakkers}}, \bibinfo {author}
  {\bibfnamefont {A.~R.}\ \bibnamefont {Akhmerov}},\ and\ \bibinfo {author}
  {\bibfnamefont {L.}~\bibnamefont {DiCarlo}},\ }\href
  {https://doi.org/10.1103/PhysRevLett.115.127002} {\bibfield  {journal}
  {\bibinfo  {journal} {Phys. Rev. Lett.}\ }\textbf {\bibinfo {volume} {115}},\
  \bibinfo {pages} {127002} (\bibinfo {year} {2015})}\BibitemShut {NoStop}%
\bibitem [{\citenamefont {Larsen}\ \emph {et~al.}(2015)\citenamefont {Larsen},
  \citenamefont {Petersson}, \citenamefont {Kuemmeth}, \citenamefont
  {Jespersen}, \citenamefont {Krogstrup}, \citenamefont {Nyg\aa{}rd},\ and\
  \citenamefont {Marcus}}]{Larsen2015}%
  \BibitemOpen
  \bibfield  {author} {\bibinfo {author} {\bibfnamefont {T.~W.}\ \bibnamefont
  {Larsen}}, \bibinfo {author} {\bibfnamefont {K.~D.}\ \bibnamefont
  {Petersson}}, \bibinfo {author} {\bibfnamefont {F.}~\bibnamefont {Kuemmeth}},
  \bibinfo {author} {\bibfnamefont {T.~S.}\ \bibnamefont {Jespersen}}, \bibinfo
  {author} {\bibfnamefont {P.}~\bibnamefont {Krogstrup}}, \bibinfo {author}
  {\bibfnamefont {J.}~\bibnamefont {Nyg\aa{}rd}},\ and\ \bibinfo {author}
  {\bibfnamefont {C.~M.}\ \bibnamefont {Marcus}},\ }\href
  {https://doi.org/10.1103/PhysRevLett.115.127001} {\bibfield  {journal}
  {\bibinfo  {journal} {Phys. Rev. Lett.}\ }\textbf {\bibinfo {volume} {115}},\
  \bibinfo {pages} {127001} (\bibinfo {year} {2015})}\BibitemShut {NoStop}%
\bibitem [{\citenamefont {Tosi}\ \emph {et~al.}(2019)\citenamefont {Tosi},
  \citenamefont {Metzger}, \citenamefont {Goffman}, \citenamefont {Urbina},
  \citenamefont {Pothier}, \citenamefont {Park}, \citenamefont {Yeyati},
  \citenamefont {Nyg\aa{}rd},\ and\ \citenamefont {Krogstrup}}]{Tosi2019}%
  \BibitemOpen
  \bibfield  {author} {\bibinfo {author} {\bibfnamefont {L.}~\bibnamefont
  {Tosi}}, \bibinfo {author} {\bibfnamefont {C.}~\bibnamefont {Metzger}},
  \bibinfo {author} {\bibfnamefont {M.~F.}\ \bibnamefont {Goffman}}, \bibinfo
  {author} {\bibfnamefont {C.}~\bibnamefont {Urbina}}, \bibinfo {author}
  {\bibfnamefont {H.}~\bibnamefont {Pothier}}, \bibinfo {author} {\bibfnamefont
  {S.}~\bibnamefont {Park}}, \bibinfo {author} {\bibfnamefont {A.~L.}\
  \bibnamefont {Yeyati}}, \bibinfo {author} {\bibfnamefont {J.}~\bibnamefont
  {Nyg\aa{}rd}},\ and\ \bibinfo {author} {\bibfnamefont {P.}~\bibnamefont
  {Krogstrup}},\ }\href {https://doi.org/10.1103/PhysRevX.9.011010} {\bibfield
  {journal} {\bibinfo  {journal} {Phys. Rev. X}\ }\textbf {\bibinfo {volume}
  {9}},\ \bibinfo {pages} {011010} (\bibinfo {year} {2019})}\BibitemShut
  {NoStop}%
\bibitem [{\citenamefont {Metzger}\ \emph {et~al.}(2021)\citenamefont
  {Metzger}, \citenamefont {Park}, \citenamefont {Tosi}, \citenamefont
  {Janvier}, \citenamefont {Reynoso}, \citenamefont {Goffman}, \citenamefont
  {Urbina}, \citenamefont {Levy~Yeyati},\ and\ \citenamefont
  {Pothier}}]{Metzger2021}%
  \BibitemOpen
  \bibfield  {author} {\bibinfo {author} {\bibfnamefont {C.}~\bibnamefont
  {Metzger}}, \bibinfo {author} {\bibfnamefont {S.}~\bibnamefont {Park}},
  \bibinfo {author} {\bibfnamefont {L.}~\bibnamefont {Tosi}}, \bibinfo {author}
  {\bibfnamefont {C.}~\bibnamefont {Janvier}}, \bibinfo {author} {\bibfnamefont
  {A.~A.}\ \bibnamefont {Reynoso}}, \bibinfo {author} {\bibfnamefont {M.~F.}\
  \bibnamefont {Goffman}}, \bibinfo {author} {\bibfnamefont {C.}~\bibnamefont
  {Urbina}}, \bibinfo {author} {\bibfnamefont {A.}~\bibnamefont
  {Levy~Yeyati}},\ and\ \bibinfo {author} {\bibfnamefont {H.}~\bibnamefont
  {Pothier}},\ }\href {https://doi.org/10.1103/PhysRevResearch.3.013036}
  {\bibfield  {journal} {\bibinfo  {journal} {Phys. Rev. Research}\ }\textbf
  {\bibinfo {volume} {3}},\ \bibinfo {pages} {013036} (\bibinfo {year}
  {2021})}\BibitemShut {NoStop}%
\bibitem [{\citenamefont {Zellekens}\ \emph {et~al.}(2022)\citenamefont
  {Zellekens}, \citenamefont {Deacon}, \citenamefont {Perla}, \citenamefont
  {Gr{\"u}tzmacher}, \citenamefont {Lepsa}, \citenamefont {Sch{\"a}pers},\ and\
  \citenamefont {Ishibashi}}]{Zellekens2022}%
  \BibitemOpen
  \bibfield  {author} {\bibinfo {author} {\bibfnamefont {P.}~\bibnamefont
  {Zellekens}}, \bibinfo {author} {\bibfnamefont {R.}~\bibnamefont {Deacon}},
  \bibinfo {author} {\bibfnamefont {P.}~\bibnamefont {Perla}}, \bibinfo
  {author} {\bibfnamefont {D.}~\bibnamefont {Gr{\"u}tzmacher}}, \bibinfo
  {author} {\bibfnamefont {M.~I.}\ \bibnamefont {Lepsa}}, \bibinfo {author}
  {\bibfnamefont {T.}~\bibnamefont {Sch{\"a}pers}},\ and\ \bibinfo {author}
  {\bibfnamefont {K.}~\bibnamefont {Ishibashi}},\ }\href
  {https://doi.org/https://doi.org/10.1038/s42005-022-01035-6} {\bibfield
  {journal} {\bibinfo  {journal} {Communications Physics}\ }\textbf {\bibinfo
  {volume} {5}},\ \bibinfo {pages} {267} (\bibinfo {year} {2022})}\BibitemShut
  {NoStop}%
\bibitem [{\citenamefont {Yazdani}\ \emph {et~al.}(2023)\citenamefont
  {Yazdani}, \citenamefont {von Oppen}, \citenamefont {Halperin},\ and\
  \citenamefont {Yacoby}}]{Yazdani2023}%
  \BibitemOpen
  \bibfield  {author} {\bibinfo {author} {\bibfnamefont {A.}~\bibnamefont
  {Yazdani}}, \bibinfo {author} {\bibfnamefont {F.}~\bibnamefont {von Oppen}},
  \bibinfo {author} {\bibfnamefont {B.~I.}\ \bibnamefont {Halperin}},\ and\
  \bibinfo {author} {\bibfnamefont {A.}~\bibnamefont {Yacoby}},\ }\href
  {https://doi.org/10.1126/science.ade0850} {\bibfield  {journal} {\bibinfo
  {journal} {Science}\ }\textbf {\bibinfo {volume} {380}},\ \bibinfo {pages}
  {eade0850} (\bibinfo {year} {2023})}\BibitemShut {NoStop}%
\bibitem [{\citenamefont {Wirths}\ \emph {et~al.}(2011)\citenamefont {Wirths},
  \citenamefont {Weis}, \citenamefont {Winden}, \citenamefont {Sladek},
  \citenamefont {Volk}, \citenamefont {Alagha}, \citenamefont {Weirich},
  \citenamefont {von~der Ahe}, \citenamefont {Hardtdegen}, \citenamefont
  {L\"{u}th}, \citenamefont {Demarina}, \citenamefont {Gr\"{u}tzmacher},\ and\
  \citenamefont {Sch\"{a}pers}}]{Wirths2011}%
  \BibitemOpen
  \bibfield  {author} {\bibinfo {author} {\bibfnamefont {S.}~\bibnamefont
  {Wirths}}, \bibinfo {author} {\bibfnamefont {K.}~\bibnamefont {Weis}},
  \bibinfo {author} {\bibfnamefont {A.}~\bibnamefont {Winden}}, \bibinfo
  {author} {\bibfnamefont {K.}~\bibnamefont {Sladek}}, \bibinfo {author}
  {\bibfnamefont {C.}~\bibnamefont {Volk}}, \bibinfo {author} {\bibfnamefont
  {S.}~\bibnamefont {Alagha}}, \bibinfo {author} {\bibfnamefont {T.~E.}\
  \bibnamefont {Weirich}}, \bibinfo {author} {\bibfnamefont {M.}~\bibnamefont
  {von~der Ahe}}, \bibinfo {author} {\bibfnamefont {H.}~\bibnamefont
  {Hardtdegen}}, \bibinfo {author} {\bibfnamefont {H.}~\bibnamefont
  {L\"{u}th}}, \bibinfo {author} {\bibfnamefont {N.}~\bibnamefont {Demarina}},
  \bibinfo {author} {\bibfnamefont {D.}~\bibnamefont {Gr\"{u}tzmacher}},\ and\
  \bibinfo {author} {\bibfnamefont {T.}~\bibnamefont {Sch\"{a}pers}},\ }\href
  {https://doi.org/10.1063/1.3631026} {\bibfield  {journal} {\bibinfo
  {journal} {Journal of Applied Physics}\ }\textbf {\bibinfo {volume} {110}},\
  \bibinfo {eid} {053709} (\bibinfo {year} {2011})}\BibitemShut {NoStop}%
\bibitem [{\citenamefont {Tserkovnyak}\ and\ \citenamefont
  {Halperin}(2006)}]{Tserkovnyak2006}%
  \BibitemOpen
  \bibfield  {author} {\bibinfo {author} {\bibfnamefont {Y.}~\bibnamefont
  {Tserkovnyak}}\ and\ \bibinfo {author} {\bibfnamefont {B.~I.}\ \bibnamefont
  {Halperin}},\ }\href {https://doi.org/10.1103/PhysRevB.74.245327} {\bibfield
  {journal} {\bibinfo  {journal} {Physical Review B (Condensed Matter and
  Materials Physics)}\ }\textbf {\bibinfo {volume} {74}},\ \bibinfo {eid}
  {245327} (\bibinfo {year} {2006})}\BibitemShut {NoStop}%
\bibitem [{\citenamefont {Rosdahl}\ \emph {et~al.}(2014)\citenamefont
  {Rosdahl}, \citenamefont {Manolescu},\ and\ \citenamefont
  {Gudmundsson}}]{Rosdahl2014}%
  \BibitemOpen
  \bibfield  {author} {\bibinfo {author} {\bibfnamefont {T.~O.}\ \bibnamefont
  {Rosdahl}}, \bibinfo {author} {\bibfnamefont {A.}~\bibnamefont {Manolescu}},\
  and\ \bibinfo {author} {\bibfnamefont {V.}~\bibnamefont {Gudmundsson}},\
  }\href {https://doi.org/10.1103/PhysRevB.90.035421} {\bibfield  {journal}
  {\bibinfo  {journal} {Phys. Rev. B}\ }\textbf {\bibinfo {volume} {90}},\
  \bibinfo {pages} {035421} (\bibinfo {year} {2014})}\BibitemShut {NoStop}%
\bibitem [{\citenamefont {Ferrari}\ \emph {et~al.}(2008)\citenamefont
  {Ferrari}, \citenamefont {Bertoni}, \citenamefont {Goldoni},\ and\
  \citenamefont {Molinari}}]{Ferrari2008}%
  \BibitemOpen
  \bibfield  {author} {\bibinfo {author} {\bibfnamefont {G.}~\bibnamefont
  {Ferrari}}, \bibinfo {author} {\bibfnamefont {A.}~\bibnamefont {Bertoni}},
  \bibinfo {author} {\bibfnamefont {G.}~\bibnamefont {Goldoni}},\ and\ \bibinfo
  {author} {\bibfnamefont {E.}~\bibnamefont {Molinari}},\ }\href
  {https://doi.org/10.1103/PhysRevB.78.115326} {\bibfield  {journal} {\bibinfo
  {journal} {Physical Review B (Condensed Matter and Materials Physics)}\
  }\textbf {\bibinfo {volume} {78}},\ \bibinfo {eid} {115326} (\bibinfo {year}
  {2008})}\BibitemShut {NoStop}%
\bibitem [{\citenamefont {Manolescu}\ \emph {et~al.}(2016)\citenamefont
  {Manolescu}, \citenamefont {Nemnes}, \citenamefont {Sitek}, \citenamefont
  {Rosdahl}, \citenamefont {Erlingsson},\ and\ \citenamefont
  {Gudmundsson}}]{Manolescu2016}%
  \BibitemOpen
  \bibfield  {author} {\bibinfo {author} {\bibfnamefont {A.}~\bibnamefont
  {Manolescu}}, \bibinfo {author} {\bibfnamefont {G.~A.}\ \bibnamefont
  {Nemnes}}, \bibinfo {author} {\bibfnamefont {A.}~\bibnamefont {Sitek}},
  \bibinfo {author} {\bibfnamefont {T.~O.}\ \bibnamefont {Rosdahl}}, \bibinfo
  {author} {\bibfnamefont {S.~I.}\ \bibnamefont {Erlingsson}},\ and\ \bibinfo
  {author} {\bibfnamefont {V.}~\bibnamefont {Gudmundsson}},\ }\href
  {https://doi.org/10.1103/PhysRevB.93.205445} {\bibfield  {journal} {\bibinfo
  {journal} {Phys. Rev. B}\ }\textbf {\bibinfo {volume} {93}},\ \bibinfo
  {pages} {205445} (\bibinfo {year} {2016})}\BibitemShut {NoStop}%
\bibitem [{\citenamefont {Aharonov}\ and\ \citenamefont
  {Bohm}(1959)}]{Aharonov1959}%
  \BibitemOpen
  \bibfield  {author} {\bibinfo {author} {\bibfnamefont {Y.}~\bibnamefont
  {Aharonov}}\ and\ \bibinfo {author} {\bibfnamefont {D.}~\bibnamefont
  {Bohm}},\ }\href {https://doi.org/10.1103/PhysRev.115.485} {\bibfield
  {journal} {\bibinfo  {journal} {Phys. Rev.}\ }\textbf {\bibinfo {volume}
  {115}},\ \bibinfo {pages} {485} (\bibinfo {year} {1959})}\BibitemShut
  {NoStop}%
\bibitem [{\citenamefont {G\"ul}\ \emph
  {et~al.}(2014{\natexlab{a}})\citenamefont {G\"ul}, \citenamefont {Demarina},
  \citenamefont {Bl\"omers}, \citenamefont {Rieger}, \citenamefont {L\"uth},
  \citenamefont {Lepsa}, \citenamefont {Gr\"utzmacher},\ and\ \citenamefont
  {Sch\"apers}}]{Guel2014}%
  \BibitemOpen
  \bibfield  {author} {\bibinfo {author} {\bibfnamefont {O.}~\bibnamefont
  {G\"ul}}, \bibinfo {author} {\bibfnamefont {N.}~\bibnamefont {Demarina}},
  \bibinfo {author} {\bibfnamefont {C.}~\bibnamefont {Bl\"omers}}, \bibinfo
  {author} {\bibfnamefont {T.}~\bibnamefont {Rieger}}, \bibinfo {author}
  {\bibfnamefont {H.}~\bibnamefont {L\"uth}}, \bibinfo {author} {\bibfnamefont
  {M.~I.}\ \bibnamefont {Lepsa}}, \bibinfo {author} {\bibfnamefont
  {D.}~\bibnamefont {Gr\"utzmacher}},\ and\ \bibinfo {author} {\bibfnamefont
  {T.}~\bibnamefont {Sch\"apers}},\ }\href
  {https://doi.org/10.1103/PhysRevB.89.045417} {\bibfield  {journal} {\bibinfo
  {journal} {Phys. Rev. B}\ }\textbf {\bibinfo {volume} {89}},\ \bibinfo
  {pages} {045417} (\bibinfo {year} {2014}{\natexlab{a}})}\BibitemShut
  {NoStop}%
\bibitem [{\citenamefont {Bl\"omers}\ \emph {et~al.}(2013)\citenamefont
  {Bl\"omers}, \citenamefont {Rieger}, \citenamefont {Zellekens}, \citenamefont
  {Haas}, \citenamefont {Lepsa}, \citenamefont {.Hardtdegen}, \citenamefont
  {G\"ul}, \citenamefont {Demarina}, \citenamefont {Gr\"utzmacher},
  \citenamefont {L\"uth},\ and\ \citenamefont {Sch\"apers}}]{Bloemers2013}%
  \BibitemOpen
  \bibfield  {author} {\bibinfo {author} {\bibfnamefont {C.}~\bibnamefont
  {Bl\"omers}}, \bibinfo {author} {\bibfnamefont {T.}~\bibnamefont {Rieger}},
  \bibinfo {author} {\bibfnamefont {P.}~\bibnamefont {Zellekens}}, \bibinfo
  {author} {\bibfnamefont {F.}~\bibnamefont {Haas}}, \bibinfo {author}
  {\bibfnamefont {M.~I.}\ \bibnamefont {Lepsa}}, \bibinfo {author}
  {\bibfnamefont {H.}~\bibnamefont {.Hardtdegen}}, \bibinfo {author}
  {\bibfnamefont {O.}~\bibnamefont {G\"ul}}, \bibinfo {author} {\bibfnamefont
  {N.}~\bibnamefont {Demarina}}, \bibinfo {author} {\bibfnamefont
  {D.}~\bibnamefont {Gr\"utzmacher}}, \bibinfo {author} {\bibfnamefont
  {H.}~\bibnamefont {L\"uth}},\ and\ \bibinfo {author} {\bibfnamefont
  {T.}~\bibnamefont {Sch\"apers}},\ }\href
  {https://doi.org/10.1088/0957-4484/24/3/035203} {\bibfield  {journal}
  {\bibinfo  {journal} {Nanotechnology}\ }\textbf {\bibinfo {volume} {24}},\
  \bibinfo {pages} {035203} (\bibinfo {year} {2013})}\BibitemShut {NoStop}%
\bibitem [{\citenamefont {Haas}\ \emph {et~al.}(2016)\citenamefont {Haas},
  \citenamefont {Wenz}, \citenamefont {Zellekens}, \citenamefont {Demarina},
  \citenamefont {Rieger}, \citenamefont {Lepsa}, \citenamefont
  {Gr{\"u}tzmacher}, \citenamefont {L{\"u}th},\ and\ \citenamefont
  {Sch{\"a}pers}}]{Haas2016}%
  \BibitemOpen
  \bibfield  {author} {\bibinfo {author} {\bibfnamefont {F.}~\bibnamefont
  {Haas}}, \bibinfo {author} {\bibfnamefont {T.}~\bibnamefont {Wenz}}, \bibinfo
  {author} {\bibfnamefont {P.}~\bibnamefont {Zellekens}}, \bibinfo {author}
  {\bibfnamefont {N.}~\bibnamefont {Demarina}}, \bibinfo {author}
  {\bibfnamefont {T.}~\bibnamefont {Rieger}}, \bibinfo {author} {\bibfnamefont
  {M.}~\bibnamefont {Lepsa}}, \bibinfo {author} {\bibfnamefont
  {D.}~\bibnamefont {Gr{\"u}tzmacher}}, \bibinfo {author} {\bibfnamefont
  {H.}~\bibnamefont {L{\"u}th}},\ and\ \bibinfo {author} {\bibfnamefont
  {T.}~\bibnamefont {Sch{\"a}pers}},\ }\href
  {https://doi.org/10.1038/srep24573} {\bibfield  {journal} {\bibinfo
  {journal} {Scientific Reports}\ }\textbf {\bibinfo {volume} {6}},\ \bibinfo
  {pages} {24573} (\bibinfo {year} {2016})}\BibitemShut {NoStop}%
\bibitem [{\citenamefont {Haas}\ \emph {et~al.}(2017)\citenamefont {Haas},
  \citenamefont {Zellekens}, \citenamefont {Wenz}, \citenamefont {Demarina},
  \citenamefont {Rieger}, \citenamefont {Lepsa}, \citenamefont {Grützmacher},
  \citenamefont {Lüth},\ and\ \citenamefont {Schäpers}}]{Haas2017}%
  \BibitemOpen
  \bibfield  {author} {\bibinfo {author} {\bibfnamefont {F.}~\bibnamefont
  {Haas}}, \bibinfo {author} {\bibfnamefont {P.}~\bibnamefont {Zellekens}},
  \bibinfo {author} {\bibfnamefont {T.}~\bibnamefont {Wenz}}, \bibinfo {author}
  {\bibfnamefont {N.}~\bibnamefont {Demarina}}, \bibinfo {author}
  {\bibfnamefont {T.}~\bibnamefont {Rieger}}, \bibinfo {author} {\bibfnamefont
  {M.~I.}\ \bibnamefont {Lepsa}}, \bibinfo {author} {\bibfnamefont
  {D.}~\bibnamefont {Grützmacher}}, \bibinfo {author} {\bibfnamefont
  {H.}~\bibnamefont {Lüth}},\ and\ \bibinfo {author} {\bibfnamefont
  {T.}~\bibnamefont {Schäpers}},\ }\href
  {https://doi.org/https://doi.org/10.1088/1361-6528/aa887d} {\bibfield
  {journal} {\bibinfo  {journal} {Nanotechnology}\ }\textbf {\bibinfo {volume}
  {28}},\ \bibinfo {pages} {445202} (\bibinfo {year} {2017})}\BibitemShut
  {NoStop}%
\bibitem [{\citenamefont {Zellekens}\ \emph {et~al.}(2020)\citenamefont
  {Zellekens}, \citenamefont {Demarina}, \citenamefont {Jan{\ss}en},
  \citenamefont {Rieger}, \citenamefont {Lepsa}, \citenamefont {Perla},
  \citenamefont {Panaitov}, \citenamefont {Lüth}, \citenamefont
  {Grützmacher},\ and\ \citenamefont {Schäpers}}]{Zellekens2020}%
  \BibitemOpen
  \bibfield  {author} {\bibinfo {author} {\bibfnamefont {P.}~\bibnamefont
  {Zellekens}}, \bibinfo {author} {\bibfnamefont {N.}~\bibnamefont {Demarina}},
  \bibinfo {author} {\bibfnamefont {J.}~\bibnamefont {Jan{\ss}en}}, \bibinfo
  {author} {\bibfnamefont {T.}~\bibnamefont {Rieger}}, \bibinfo {author}
  {\bibfnamefont {M.~I.}\ \bibnamefont {Lepsa}}, \bibinfo {author}
  {\bibfnamefont {P.}~\bibnamefont {Perla}}, \bibinfo {author} {\bibfnamefont
  {G.}~\bibnamefont {Panaitov}}, \bibinfo {author} {\bibfnamefont
  {H.}~\bibnamefont {Lüth}}, \bibinfo {author} {\bibfnamefont
  {D.}~\bibnamefont {Grützmacher}},\ and\ \bibinfo {author} {\bibfnamefont
  {T.}~\bibnamefont {Schäpers}},\ }\href
  {https://doi.org/10.1088/1361-6641/ab8396} {\bibfield  {journal} {\bibinfo
  {journal} {Semiconductor Science and Technology}\ }\textbf {\bibinfo {volume}
  {35}},\ \bibinfo {pages} {085003} (\bibinfo {year} {2020})}\BibitemShut
  {NoStop}%
\bibitem [{\citenamefont {Jung}\ \emph {et~al.}(2008)\citenamefont {Jung},
  \citenamefont {Lee}, \citenamefont {Song}, \citenamefont {Kim}, \citenamefont
  {Lee}, \citenamefont {Kim}, \citenamefont {Park}, \citenamefont {Choi},
  \citenamefont {Katsumoto}, \citenamefont {Lee},\ and\ \citenamefont
  {Kim}}]{Jung2008}%
  \BibitemOpen
  \bibfield  {author} {\bibinfo {author} {\bibfnamefont {M.}~\bibnamefont
  {Jung}}, \bibinfo {author} {\bibfnamefont {J.~S.}\ \bibnamefont {Lee}},
  \bibinfo {author} {\bibfnamefont {W.}~\bibnamefont {Song}}, \bibinfo {author}
  {\bibfnamefont {Y.~H.}\ \bibnamefont {Kim}}, \bibinfo {author} {\bibfnamefont
  {S.~D.}\ \bibnamefont {Lee}}, \bibinfo {author} {\bibfnamefont
  {N.}~\bibnamefont {Kim}}, \bibinfo {author} {\bibfnamefont {J.}~\bibnamefont
  {Park}}, \bibinfo {author} {\bibfnamefont {M.-S.}\ \bibnamefont {Choi}},
  \bibinfo {author} {\bibfnamefont {S.}~\bibnamefont {Katsumoto}}, \bibinfo
  {author} {\bibfnamefont {H.}~\bibnamefont {Lee}},\ and\ \bibinfo {author}
  {\bibfnamefont {J.}~\bibnamefont {Kim}},\ }\href
  {https://doi.org/10.1021/nl801506w} {\bibfield  {journal} {\bibinfo
  {journal} {Nano Letters}\ }\textbf {\bibinfo {volume} {8}},\ \bibinfo {pages}
  {3189} (\bibinfo {year} {2008})}\BibitemShut {NoStop}%
\bibitem [{\citenamefont {Peng}\ \emph {et~al.}(2010)\citenamefont {Peng},
  \citenamefont {Lai}, \citenamefont {Kong}, \citenamefont {Meister},
  \citenamefont {Chen}, \citenamefont {Qi}, \citenamefont {Zhang},
  \citenamefont {Shen},\ and\ \citenamefont {Cui}}]{Peng2010}%
  \BibitemOpen
  \bibfield  {author} {\bibinfo {author} {\bibfnamefont {H.}~\bibnamefont
  {Peng}}, \bibinfo {author} {\bibfnamefont {K.}~\bibnamefont {Lai}}, \bibinfo
  {author} {\bibfnamefont {D.}~\bibnamefont {Kong}}, \bibinfo {author}
  {\bibfnamefont {S.}~\bibnamefont {Meister}}, \bibinfo {author} {\bibfnamefont
  {Y.}~\bibnamefont {Chen}}, \bibinfo {author} {\bibfnamefont {X.-L.}\
  \bibnamefont {Qi}}, \bibinfo {author} {\bibfnamefont {S.-C.}\ \bibnamefont
  {Zhang}}, \bibinfo {author} {\bibfnamefont {Z.-X.}\ \bibnamefont {Shen}},\
  and\ \bibinfo {author} {\bibfnamefont {Y.}~\bibnamefont {Cui}},\ }\href
  {https://doi.org/10.1038/nmat2609} {\bibfield  {journal} {\bibinfo  {journal}
  {Nat. Materials}\ }\textbf {\bibinfo {volume} {9}},\ \bibinfo {pages} {225}
  (\bibinfo {year} {2010})}\BibitemShut {NoStop}%
\bibitem [{\citenamefont {Cho}\ \emph {et~al.}(2015)\citenamefont {Cho},
  \citenamefont {Dellabetta}, \citenamefont {Zhong}, \citenamefont
  {Schneeloch}, \citenamefont {Liu}, \citenamefont {Gu}, \citenamefont
  {Gilbert},\ and\ \citenamefont {Mason}}]{Cho2015}%
  \BibitemOpen
  \bibfield  {author} {\bibinfo {author} {\bibfnamefont {S.}~\bibnamefont
  {Cho}}, \bibinfo {author} {\bibfnamefont {B.}~\bibnamefont {Dellabetta}},
  \bibinfo {author} {\bibfnamefont {R.}~\bibnamefont {Zhong}}, \bibinfo
  {author} {\bibfnamefont {J.}~\bibnamefont {Schneeloch}}, \bibinfo {author}
  {\bibfnamefont {T.}~\bibnamefont {Liu}}, \bibinfo {author} {\bibfnamefont
  {G.}~\bibnamefont {Gu}}, \bibinfo {author} {\bibfnamefont {M.~J.}\
  \bibnamefont {Gilbert}},\ and\ \bibinfo {author} {\bibfnamefont
  {N.}~\bibnamefont {Mason}},\ }\href {https://doi.org/10.1038/ncomms8634}
  {\bibfield  {journal} {\bibinfo  {journal} {Nature Communications}\ }\textbf
  {\bibinfo {volume} {6}},\ \bibinfo {pages} {7634} (\bibinfo {year}
  {2015})}\BibitemShut {NoStop}%
\bibitem [{\citenamefont {Arango}\ \emph {et~al.}(2016)\citenamefont {Arango},
  \citenamefont {Huang}, \citenamefont {Chen}, \citenamefont {Avila},
  \citenamefont {Asensio}, \citenamefont {Gr\"utzmacher}, \citenamefont
  {L\"uth}, \citenamefont {Lu},\ and\ \citenamefont {Sch\"apers}}]{Arango2016}%
  \BibitemOpen
  \bibfield  {author} {\bibinfo {author} {\bibfnamefont {Y.~C.}\ \bibnamefont
  {Arango}}, \bibinfo {author} {\bibfnamefont {L.}~\bibnamefont {Huang}},
  \bibinfo {author} {\bibfnamefont {C.}~\bibnamefont {Chen}}, \bibinfo {author}
  {\bibfnamefont {J.}~\bibnamefont {Avila}}, \bibinfo {author} {\bibfnamefont
  {M.~C.}\ \bibnamefont {Asensio}}, \bibinfo {author} {\bibfnamefont
  {D.}~\bibnamefont {Gr\"utzmacher}}, \bibinfo {author} {\bibfnamefont
  {H.}~\bibnamefont {L\"uth}}, \bibinfo {author} {\bibfnamefont {J.~G.}\
  \bibnamefont {Lu}},\ and\ \bibinfo {author} {\bibfnamefont {T.}~\bibnamefont
  {Sch\"apers}},\ }\href {https://doi.org/10.1038/srep29493} {\bibfield
  {journal} {\bibinfo  {journal} {Scientific Reports}\ }\textbf {\bibinfo
  {volume} {6}},\ \bibinfo {pages} {29493} (\bibinfo {year}
  {2016})}\BibitemShut {NoStop}%
\bibitem [{\citenamefont {Ziegler}\ \emph {et~al.}(2018)\citenamefont
  {Ziegler}, \citenamefont {Kozlovsky}, \citenamefont {Gorini}, \citenamefont
  {Liu}, \citenamefont {Weish\"aupl}, \citenamefont {Maier}, \citenamefont
  {Fischer}, \citenamefont {Kozlov}, \citenamefont {Kvon}, \citenamefont
  {Mikhailov}, \citenamefont {Dvoretsky}, \citenamefont {Richter},\ and\
  \citenamefont {Weiss}}]{Ziegler2018}%
  \BibitemOpen
  \bibfield  {author} {\bibinfo {author} {\bibfnamefont {J.}~\bibnamefont
  {Ziegler}}, \bibinfo {author} {\bibfnamefont {R.}~\bibnamefont {Kozlovsky}},
  \bibinfo {author} {\bibfnamefont {C.}~\bibnamefont {Gorini}}, \bibinfo
  {author} {\bibfnamefont {M.-H.}\ \bibnamefont {Liu}}, \bibinfo {author}
  {\bibfnamefont {S.}~\bibnamefont {Weish\"aupl}}, \bibinfo {author}
  {\bibfnamefont {H.}~\bibnamefont {Maier}}, \bibinfo {author} {\bibfnamefont
  {R.}~\bibnamefont {Fischer}}, \bibinfo {author} {\bibfnamefont {D.~A.}\
  \bibnamefont {Kozlov}}, \bibinfo {author} {\bibfnamefont {Z.~D.}\
  \bibnamefont {Kvon}}, \bibinfo {author} {\bibfnamefont {N.}~\bibnamefont
  {Mikhailov}}, \bibinfo {author} {\bibfnamefont {S.~A.}\ \bibnamefont
  {Dvoretsky}}, \bibinfo {author} {\bibfnamefont {K.}~\bibnamefont {Richter}},\
  and\ \bibinfo {author} {\bibfnamefont {D.}~\bibnamefont {Weiss}},\ }\href
  {https://doi.org/10.1103/PhysRevB.97.035157} {\bibfield  {journal} {\bibinfo
  {journal} {Phys. Rev. B}\ }\textbf {\bibinfo {volume} {97}},\ \bibinfo
  {pages} {035157} (\bibinfo {year} {2018})}\BibitemShut {NoStop}%
\bibitem [{\citenamefont {Rosenbach}\ \emph {et~al.}(2022)\citenamefont
  {Rosenbach}, \citenamefont {Moors}, \citenamefont {Jalil}, \citenamefont
  {K\"olzer}, \citenamefont {Zimmermann}, \citenamefont {Schubert},
  \citenamefont {Karimzadah}, \citenamefont {Mussler}, \citenamefont
  {Sch\"uffelgen}, \citenamefont {Gr\"utzmacher}, \citenamefont {L\"uth},\ and\
  \citenamefont {Sch\"apers}}]{Rosenbach2022}%
  \BibitemOpen
  \bibfield  {author} {\bibinfo {author} {\bibfnamefont {D.}~\bibnamefont
  {Rosenbach}}, \bibinfo {author} {\bibfnamefont {K.}~\bibnamefont {Moors}},
  \bibinfo {author} {\bibfnamefont {A.~R.}\ \bibnamefont {Jalil}}, \bibinfo
  {author} {\bibfnamefont {J.}~\bibnamefont {K\"olzer}}, \bibinfo {author}
  {\bibfnamefont {E.}~\bibnamefont {Zimmermann}}, \bibinfo {author}
  {\bibfnamefont {J.}~\bibnamefont {Schubert}}, \bibinfo {author}
  {\bibfnamefont {S.}~\bibnamefont {Karimzadah}}, \bibinfo {author}
  {\bibfnamefont {G.}~\bibnamefont {Mussler}}, \bibinfo {author} {\bibfnamefont
  {P.}~\bibnamefont {Sch\"uffelgen}}, \bibinfo {author} {\bibfnamefont
  {D.}~\bibnamefont {Gr\"utzmacher}}, \bibinfo {author} {\bibfnamefont
  {H.}~\bibnamefont {L\"uth}},\ and\ \bibinfo {author} {\bibfnamefont
  {T.}~\bibnamefont {Sch\"apers}},\ }\href
  {https://doi.org/10.21468/SciPostPhysCore.5.1.017} {\bibfield  {journal}
  {\bibinfo  {journal} {SciPost Phys. Core}\ }\textbf {\bibinfo {volume} {5}},\
  \bibinfo {pages} {17} (\bibinfo {year} {2022})}\BibitemShut {NoStop}%
\bibitem [{\citenamefont {Kim}\ \emph {et~al.}(2020)\citenamefont {Kim},
  \citenamefont {Hwang}, \citenamefont {Kim}, \citenamefont {Hou},
  \citenamefont {Yu}, \citenamefont {Sim},\ and\ \citenamefont
  {Doh}}]{Kim2020}%
  \BibitemOpen
  \bibfield  {author} {\bibinfo {author} {\bibfnamefont {H.-S.}\ \bibnamefont
  {Kim}}, \bibinfo {author} {\bibfnamefont {T.-H.}\ \bibnamefont {Hwang}},
  \bibinfo {author} {\bibfnamefont {N.-H.}\ \bibnamefont {Kim}}, \bibinfo
  {author} {\bibfnamefont {Y.}~\bibnamefont {Hou}}, \bibinfo {author}
  {\bibfnamefont {D.}~\bibnamefont {Yu}}, \bibinfo {author} {\bibfnamefont
  {H.-S.}\ \bibnamefont {Sim}},\ and\ \bibinfo {author} {\bibfnamefont {Y.-J.}\
  \bibnamefont {Doh}},\ }\href
  {https://doi.org/https://doi.org/10.1021/acsnano.0c06892} {\bibfield
  {journal} {\bibinfo  {journal} {ACS Nano}\ }\textbf {\bibinfo {volume}
  {14}},\ \bibinfo {pages} {14118} (\bibinfo {year} {2020})}\BibitemShut
  {NoStop}%
\bibitem [{\citenamefont {Jansen}\ \emph {et~al.}(2020)\citenamefont {Jansen},
  \citenamefont {Perla}, \citenamefont {Kaladzhian}, \citenamefont {von~den
  Driesch}, \citenamefont {Janssen}, \citenamefont {Luysberg}, \citenamefont
  {Lepsa}, \citenamefont {Gr\"utzmacher},\ and\ \citenamefont
  {Pawlis}}]{Jansen2020}%
  \BibitemOpen
  \bibfield  {author} {\bibinfo {author} {\bibfnamefont {M.~M.}\ \bibnamefont
  {Jansen}}, \bibinfo {author} {\bibfnamefont {P.}~\bibnamefont {Perla}},
  \bibinfo {author} {\bibfnamefont {M.}~\bibnamefont {Kaladzhian}}, \bibinfo
  {author} {\bibfnamefont {N.}~\bibnamefont {von~den Driesch}}, \bibinfo
  {author} {\bibfnamefont {J.}~\bibnamefont {Janssen}}, \bibinfo {author}
  {\bibfnamefont {M.}~\bibnamefont {Luysberg}}, \bibinfo {author}
  {\bibfnamefont {M.~I.}\ \bibnamefont {Lepsa}}, \bibinfo {author}
  {\bibfnamefont {D.}~\bibnamefont {Gr\"utzmacher}},\ and\ \bibinfo {author}
  {\bibfnamefont {A.}~\bibnamefont {Pawlis}},\ }\href
  {https://doi.org/https://doi.org/10.1021/acsanm.0c02241} {\bibfield
  {journal} {\bibinfo  {journal} {ACS Applied Nano Materials}\ }\textbf
  {\bibinfo {volume} {3}},\ \bibinfo {pages} {11037} (\bibinfo {year}
  {2020})}\BibitemShut {NoStop}%
\bibitem [{\citenamefont {Groth}\ \emph {et~al.}(2014)\citenamefont {Groth},
  \citenamefont {Wimmer}, \citenamefont {Akhmerov},\ and\ \citenamefont
  {Waintal}}]{Groth2014}%
  \BibitemOpen
  \bibfield  {author} {\bibinfo {author} {\bibfnamefont {C.~W.}\ \bibnamefont
  {Groth}}, \bibinfo {author} {\bibfnamefont {M.}~\bibnamefont {Wimmer}},
  \bibinfo {author} {\bibfnamefont {A.~R.}\ \bibnamefont {Akhmerov}},\ and\
  \bibinfo {author} {\bibfnamefont {X.}~\bibnamefont {Waintal}},\ }\href
  {https://doi.org/10.1088/1367-2630/16/6/063065} {\bibfield  {journal}
  {\bibinfo  {journal} {New Journal of Physics}\ }\textbf {\bibinfo {volume}
  {16}},\ \bibinfo {pages} {063065} (\bibinfo {year} {2014})}\BibitemShut
  {NoStop}%
\bibitem [{\citenamefont {Al'tshuler}\ \emph {et~al.}(1981)\citenamefont
  {Al'tshuler}, \citenamefont {Aronov},\ and\ \citenamefont
  {Spivak}}]{Altshuler1981}%
  \BibitemOpen
  \bibfield  {author} {\bibinfo {author} {\bibfnamefont {B.~L.}\ \bibnamefont
  {Al'tshuler}}, \bibinfo {author} {\bibfnamefont {A.~G.}\ \bibnamefont
  {Aronov}},\ and\ \bibinfo {author} {\bibfnamefont {B.~Z.}\ \bibnamefont
  {Spivak}},\ }\href@noop {} {\bibfield  {journal} {\bibinfo  {journal} {Pis'ma
  Zh. Eksp. Teor. Fiz. [JETP Lett. {\bf 33}, 94 (1981)]}\ }\textbf {\bibinfo
  {volume} {33}},\ \bibinfo {pages} {101} (\bibinfo {year} {1981})}\BibitemShut
  {NoStop}%
\bibitem [{\citenamefont {Sharvin}\ and\ \citenamefont
  {Sharvin}(1981)}]{Sharvin81}%
  \BibitemOpen
  \bibfield  {author} {\bibinfo {author} {\bibfnamefont {D.~Y.}\ \bibnamefont
  {Sharvin}}\ and\ \bibinfo {author} {\bibfnamefont {Y.~V.}\ \bibnamefont
  {Sharvin}},\ }\href@noop {} {\bibfield  {journal} {\bibinfo  {journal}
  {Pis'ma Zh. Eksp. Teor. Fiz.[JETP Lett. {\bf 34}, 272 (1981)]}\ }\textbf
  {\bibinfo {volume} {34}},\ \bibinfo {pages} {285} (\bibinfo {year}
  {1981})}\BibitemShut {NoStop}%
\bibitem [{\citenamefont {Aronov}\ and\ \citenamefont
  {Sharvin}(1987)}]{Aronov1987}%
  \BibitemOpen
  \bibfield  {author} {\bibinfo {author} {\bibfnamefont {A.~G.}\ \bibnamefont
  {Aronov}}\ and\ \bibinfo {author} {\bibfnamefont {Y.~V.}\ \bibnamefont
  {Sharvin}},\ }\href {https://doi.org/10.1103/RevModPhys.59.755} {\bibfield
  {journal} {\bibinfo  {journal} {Rev. Mod. Phys.}\ }\textbf {\bibinfo {volume}
  {59}},\ \bibinfo {pages} {755} (\bibinfo {year} {1987})}\BibitemShut
  {NoStop}%
\bibitem [{\citenamefont {Est\'evez~Hern\'andez}\ \emph
  {et~al.}(2010)\citenamefont {Est\'evez~Hern\'andez}, \citenamefont {Akabori},
  \citenamefont {Sladek}, \citenamefont {Volk}, \citenamefont {Alagha},
  \citenamefont {Hardtdegen}, \citenamefont {Pala}, \citenamefont {Demarina},
  \citenamefont {Gr\"utzmacher},\ and\ \citenamefont
  {Sch\"apers}}]{Estevez2010}%
  \BibitemOpen
  \bibfield  {author} {\bibinfo {author} {\bibfnamefont {S.}~\bibnamefont
  {Est\'evez~Hern\'andez}}, \bibinfo {author} {\bibfnamefont {M.}~\bibnamefont
  {Akabori}}, \bibinfo {author} {\bibfnamefont {K.}~\bibnamefont {Sladek}},
  \bibinfo {author} {\bibfnamefont {C.}~\bibnamefont {Volk}}, \bibinfo {author}
  {\bibfnamefont {S.}~\bibnamefont {Alagha}}, \bibinfo {author} {\bibfnamefont
  {H.}~\bibnamefont {Hardtdegen}}, \bibinfo {author} {\bibfnamefont {M.~G.}\
  \bibnamefont {Pala}}, \bibinfo {author} {\bibfnamefont {N.}~\bibnamefont
  {Demarina}}, \bibinfo {author} {\bibfnamefont {D.}~\bibnamefont
  {Gr\"utzmacher}},\ and\ \bibinfo {author} {\bibfnamefont {T.}~\bibnamefont
  {Sch\"apers}},\ }\href {https://doi.org/10.1103/PhysRevB.82.235303}
  {\bibfield  {journal} {\bibinfo  {journal} {Phys. Rev. B}\ }\textbf {\bibinfo
  {volume} {82}},\ \bibinfo {pages} {235303} (\bibinfo {year}
  {2010})}\BibitemShut {NoStop}%
\bibitem [{\citenamefont {Mur}\ \emph {et~al.}(2008)\citenamefont {Mur},
  \citenamefont {Harmans},\ and\ \citenamefont {van~der Wiel}}]{Mur2008}%
  \BibitemOpen
  \bibfield  {author} {\bibinfo {author} {\bibfnamefont {L.~C.}\ \bibnamefont
  {Mur}}, \bibinfo {author} {\bibfnamefont {C.~J. P.~M.}\ \bibnamefont
  {Harmans}},\ and\ \bibinfo {author} {\bibfnamefont {W.~G.}\ \bibnamefont
  {van~der Wiel}},\ }\href {https://doi.org/10.1088/1367-2630/10/7/073031}
  {\bibfield  {journal} {\bibinfo  {journal} {New Journal of Physics}\ }\textbf
  {\bibinfo {volume} {10}},\ \bibinfo {pages} {073031} (\bibinfo {year}
  {2008})}\BibitemShut {NoStop}%
\bibitem [{\citenamefont {Ren}\ \emph {et~al.}(2015)\citenamefont {Ren},
  \citenamefont {Heremans}, \citenamefont {Vijeyaragunathan}, \citenamefont
  {Mishima},\ and\ \citenamefont {Santos}}]{SLRen2015}%
  \BibitemOpen
  \bibfield  {author} {\bibinfo {author} {\bibfnamefont {S.~L.}\ \bibnamefont
  {Ren}}, \bibinfo {author} {\bibfnamefont {J.~J.}\ \bibnamefont {Heremans}},
  \bibinfo {author} {\bibfnamefont {S.}~\bibnamefont {Vijeyaragunathan}},
  \bibinfo {author} {\bibfnamefont {T.~D.}\ \bibnamefont {Mishima}},\ and\
  \bibinfo {author} {\bibfnamefont {M.~B.}\ \bibnamefont {Santos}},\ }\href
  {https://doi.org/10.1088/0953-8984/27/18/185801} {\bibfield  {journal}
  {\bibinfo  {journal} {IOP Publishing}\ }\textbf {\bibinfo {volume} {27}},\
  \bibinfo {pages} {185801} (\bibinfo {year} {2015})}\BibitemShut {NoStop}%
\bibitem [{\citenamefont {Umbach}\ \emph {et~al.}(1986)\citenamefont {Umbach},
  \citenamefont {Van~Haesendonck}, \citenamefont {Laibowitz}, \citenamefont
  {Washburn},\ and\ \citenamefont {Webb}}]{CPUmbach}%
  \BibitemOpen
  \bibfield  {author} {\bibinfo {author} {\bibfnamefont {C.~P.}\ \bibnamefont
  {Umbach}}, \bibinfo {author} {\bibfnamefont {C.}~\bibnamefont
  {Van~Haesendonck}}, \bibinfo {author} {\bibfnamefont {R.~B.}\ \bibnamefont
  {Laibowitz}}, \bibinfo {author} {\bibfnamefont {S.}~\bibnamefont
  {Washburn}},\ and\ \bibinfo {author} {\bibfnamefont {R.~A.}\ \bibnamefont
  {Webb}},\ }\href {https://doi.org/10.1103/PhysRevLett.56.386} {\bibfield
  {journal} {\bibinfo  {journal} {Phys. Rev. Lett.}\ }\textbf {\bibinfo
  {volume} {56}},\ \bibinfo {pages} {386} (\bibinfo {year} {1986})}\BibitemShut
  {NoStop}%
\bibitem [{\citenamefont {Kubo}(1957)}]{Kubo1957}%
  \BibitemOpen
  \bibfield  {author} {\bibinfo {author} {\bibfnamefont {R.}~\bibnamefont
  {Kubo}},\ }\href {https://doi.org/10.1143/JPSJ.12.570} {\bibfield  {journal}
  {\bibinfo  {journal} {J. Phys. Soc. Jpn.}\ }\textbf {\bibinfo {volume}
  {12}},\ \bibinfo {pages} {570} (\bibinfo {year} {1957})}\BibitemShut
  {NoStop}%
\bibitem [{\citenamefont {Winkler}(2003)}]{Winkler2003}%
  \BibitemOpen
  \bibfield  {author} {\bibinfo {author} {\bibfnamefont {R.}~\bibnamefont
  {Winkler}},\ }\href {https://doi.org/https://doi.org/10.1007/b13586} {}\
  (\bibinfo  {publisher} {Springer--Verlag},\ \bibinfo {address} {Berlin,
  Heidelberg, New York},\ \bibinfo {year} {2003})\BibitemShut {NoStop}%
\bibitem [{\citenamefont {Umbach}\ \emph {et~al.}(1987)\citenamefont {Umbach},
  \citenamefont {Santhanam}, \citenamefont {van Haesendonck},\ and\
  \citenamefont {Webb}}]{Umbach1987}%
  \BibitemOpen
  \bibfield  {author} {\bibinfo {author} {\bibfnamefont {C.~P.}\ \bibnamefont
  {Umbach}}, \bibinfo {author} {\bibfnamefont {P.}~\bibnamefont {Santhanam}},
  \bibinfo {author} {\bibfnamefont {C.}~\bibnamefont {van Haesendonck}},\ and\
  \bibinfo {author} {\bibfnamefont {R.~A.}\ \bibnamefont {Webb}},\ }\href
  {https://doi.org/10.1063/1.97887} {\bibfield  {journal} {\bibinfo  {journal}
  {Appl. Phys. Lett.}\ }\textbf {\bibinfo {volume} {50}},\ \bibinfo {pages}
  {1289} (\bibinfo {year} {1987})}\BibitemShut {NoStop}%
\bibitem [{\citenamefont {Lee}\ \emph {et~al.}(1987)\citenamefont {Lee},
  \citenamefont {Stone},\ and\ \citenamefont {Fukuyama}}]{Lee1987}%
  \BibitemOpen
  \bibfield  {author} {\bibinfo {author} {\bibfnamefont {P.~A.}\ \bibnamefont
  {Lee}}, \bibinfo {author} {\bibfnamefont {A.~D.}\ \bibnamefont {Stone}},\
  and\ \bibinfo {author} {\bibfnamefont {H.}~\bibnamefont {Fukuyama}},\ }\href
  {https://doi.org/10.1103/PhysRevB.35.1039} {\bibfield  {journal} {\bibinfo
  {journal} {Phys. Rev. B}\ }\textbf {\bibinfo {volume} {35}},\ \bibinfo
  {pages} {1039} (\bibinfo {year} {1987})}\BibitemShut {NoStop}%
\bibitem [{\citenamefont {Webb}\ \emph {et~al.}(1986)\citenamefont {Webb},
  \citenamefont {Washburn}, \citenamefont {Umbach}, \citenamefont {Milliken},
  \citenamefont {Laibowitz},\ and\ \citenamefont {Benoit}}]{Webb86}%
  \BibitemOpen
  \bibfield  {author} {\bibinfo {author} {\bibfnamefont {R.}~\bibnamefont
  {Webb}}, \bibinfo {author} {\bibfnamefont {S.}~\bibnamefont {Washburn}},
  \bibinfo {author} {\bibfnamefont {C.}~\bibnamefont {Umbach}}, \bibinfo
  {author} {\bibfnamefont {F.}~\bibnamefont {Milliken}}, \bibinfo {author}
  {\bibfnamefont {R.}~\bibnamefont {Laibowitz}},\ and\ \bibinfo {author}
  {\bibfnamefont {A.}~\bibnamefont {Benoit}},\ }\href
  {https://doi.org/10.1016/0378-4371(86)90218-9} {\bibfield  {journal}
  {\bibinfo  {journal} {Physica A: Statistical Mechanics and its Applications}\
  }\textbf {\bibinfo {volume} {140}},\ \bibinfo {pages} {1039} (\bibinfo {year}
  {1986})}\BibitemShut {NoStop}%
\bibitem [{\citenamefont {Seelig}\ and\ \citenamefont
  {B\"uttiker}(2001)}]{Seelig2001}%
  \BibitemOpen
  \bibfield  {author} {\bibinfo {author} {\bibfnamefont {G.}~\bibnamefont
  {Seelig}}\ and\ \bibinfo {author} {\bibfnamefont {M.}~\bibnamefont
  {B\"uttiker}},\ }\href {https://doi.org/10.1103/PhysRevB.64.245313}
  {\bibfield  {journal} {\bibinfo  {journal} {Phys. Rev. B}\ }\textbf {\bibinfo
  {volume} {64}},\ \bibinfo {pages} {245313} (\bibinfo {year}
  {2001})}\BibitemShut {NoStop}%
\bibitem [{\citenamefont {Ludwig}\ and\ \citenamefont
  {Mirlin}(2004)}]{Ludwig2004}%
  \BibitemOpen
  \bibfield  {author} {\bibinfo {author} {\bibfnamefont {T.}~\bibnamefont
  {Ludwig}}\ and\ \bibinfo {author} {\bibfnamefont {A.~D.}\ \bibnamefont
  {Mirlin}},\ }\href {https://doi.org/10.1103/PhysRevB.69.193306} {\bibfield
  {journal} {\bibinfo  {journal} {Phys. Rev. B}\ }\textbf {\bibinfo {volume}
  {69}},\ \bibinfo {pages} {193306} (\bibinfo {year} {2004})}\BibitemShut
  {NoStop}%
\bibitem [{\citenamefont {Dufouleur}\ \emph {et~al.}(2013)\citenamefont
  {Dufouleur}, \citenamefont {Veyrat}, \citenamefont {Teichgr\"aber},
  \citenamefont {Neuhaus}, \citenamefont {Nowka}, \citenamefont {Hampel},
  \citenamefont {Cayssol}, \citenamefont {Schumann}, \citenamefont {Eichler},
  \citenamefont {Schmidt}, \citenamefont {B\"uchner},\ and\ \citenamefont
  {Giraud}}]{Dufouleur2013}%
  \BibitemOpen
  \bibfield  {author} {\bibinfo {author} {\bibfnamefont {J.}~\bibnamefont
  {Dufouleur}}, \bibinfo {author} {\bibfnamefont {L.}~\bibnamefont {Veyrat}},
  \bibinfo {author} {\bibfnamefont {A.}~\bibnamefont {Teichgr\"aber}}, \bibinfo
  {author} {\bibfnamefont {S.}~\bibnamefont {Neuhaus}}, \bibinfo {author}
  {\bibfnamefont {C.}~\bibnamefont {Nowka}}, \bibinfo {author} {\bibfnamefont
  {S.}~\bibnamefont {Hampel}}, \bibinfo {author} {\bibfnamefont
  {J.}~\bibnamefont {Cayssol}}, \bibinfo {author} {\bibfnamefont
  {J.}~\bibnamefont {Schumann}}, \bibinfo {author} {\bibfnamefont
  {B.}~\bibnamefont {Eichler}}, \bibinfo {author} {\bibfnamefont {O.~G.}\
  \bibnamefont {Schmidt}}, \bibinfo {author} {\bibfnamefont {B.}~\bibnamefont
  {B\"uchner}},\ and\ \bibinfo {author} {\bibfnamefont {R.}~\bibnamefont
  {Giraud}},\ }\href {https://doi.org/10.1103/PhysRevLett.110.186806}
  {\bibfield  {journal} {\bibinfo  {journal} {Phys. Rev. Lett.}\ }\textbf
  {\bibinfo {volume} {110}},\ \bibinfo {pages} {186806} (\bibinfo {year}
  {2013})}\BibitemShut {NoStop}%
\bibitem [{\citenamefont {Oreg}\ \emph {et~al.}(2010)\citenamefont {Oreg},
  \citenamefont {Refael},\ and\ \citenamefont {von Oppen}}]{Oreg2010}%
  \BibitemOpen
  \bibfield  {author} {\bibinfo {author} {\bibfnamefont {Y.}~\bibnamefont
  {Oreg}}, \bibinfo {author} {\bibfnamefont {G.}~\bibnamefont {Refael}},\ and\
  \bibinfo {author} {\bibfnamefont {F.}~\bibnamefont {von Oppen}},\ }\href
  {https://doi.org/10.1103/PhysRevLett.105.177002} {\bibfield  {journal}
  {\bibinfo  {journal} {Phys. Rev. Lett.}\ }\textbf {\bibinfo {volume} {105}},\
  \bibinfo {pages} {177002} (\bibinfo {year} {2010})}\BibitemShut {NoStop}%
\bibitem [{\citenamefont {Lutchyn}\ \emph {et~al.}(2010)\citenamefont
  {Lutchyn}, \citenamefont {Sau},\ and\ \citenamefont
  {Das~Sarma}}]{Lutchyn2010}%
  \BibitemOpen
  \bibfield  {author} {\bibinfo {author} {\bibfnamefont {R.~M.}\ \bibnamefont
  {Lutchyn}}, \bibinfo {author} {\bibfnamefont {J.~D.}\ \bibnamefont {Sau}},\
  and\ \bibinfo {author} {\bibfnamefont {S.}~\bibnamefont {Das~Sarma}},\ }\href
  {https://doi.org/10.1103/PhysRevLett.105.077001} {\bibfield  {journal}
  {\bibinfo  {journal} {Phys. Rev. Lett.}\ }\textbf {\bibinfo {volume} {105}},\
  \bibinfo {pages} {077001} (\bibinfo {year} {2010})}\BibitemShut {NoStop}%
\bibitem [{\citenamefont {Prada}\ \emph {et~al.}(2020)\citenamefont {Prada},
  \citenamefont {San-Jose}, \citenamefont {de~Moor}, \citenamefont {Geresdi},
  \citenamefont {Lee}, \citenamefont {Klinovaja}, \citenamefont {Loss},
  \citenamefont {Nyg{\aa}rd}, \citenamefont {Aguado},\ and\ \citenamefont
  {Kouwenhoven}}]{Prada2020}%
  \BibitemOpen
  \bibfield  {author} {\bibinfo {author} {\bibfnamefont {E.}~\bibnamefont
  {Prada}}, \bibinfo {author} {\bibfnamefont {P.}~\bibnamefont {San-Jose}},
  \bibinfo {author} {\bibfnamefont {M.~W.}\ \bibnamefont {de~Moor}}, \bibinfo
  {author} {\bibfnamefont {A.}~\bibnamefont {Geresdi}}, \bibinfo {author}
  {\bibfnamefont {E.~J.}\ \bibnamefont {Lee}}, \bibinfo {author} {\bibfnamefont
  {J.}~\bibnamefont {Klinovaja}}, \bibinfo {author} {\bibfnamefont
  {D.}~\bibnamefont {Loss}}, \bibinfo {author} {\bibfnamefont {J.}~\bibnamefont
  {Nyg{\aa}rd}}, \bibinfo {author} {\bibfnamefont {R.}~\bibnamefont {Aguado}},\
  and\ \bibinfo {author} {\bibfnamefont {L.~P.}\ \bibnamefont {Kouwenhoven}},\
  }\href {https://doi.org/https://doi.org/10.1038/s42254-020-0228-y} {\bibfield
   {journal} {\bibinfo  {journal} {Nature Reviews Physics}\ }\textbf {\bibinfo
  {volume} {2}},\ \bibinfo {pages} {575} (\bibinfo {year} {2020})}\BibitemShut
  {NoStop}%
\bibitem [{\citenamefont {Zazunov}\ \emph {et~al.}(2003)\citenamefont
  {Zazunov}, \citenamefont {S.}, \citenamefont {Shumeiko}, \citenamefont
  {Bratus}, \citenamefont {Lantz},\ and\ \citenamefont {Wendin}}]{Zazunov2003}%
  \BibitemOpen
  \bibfield  {author} {\bibinfo {author} {\bibfnamefont {A.}~\bibnamefont
  {Zazunov}}, \bibinfo {author} {\bibfnamefont {V.}~\bibnamefont {S.}},
  \bibinfo {author} {\bibnamefont {Shumeiko}}, \bibinfo {author} {\bibfnamefont
  {E.~N.}\ \bibnamefont {Bratus}}, \bibinfo {author} {\bibfnamefont
  {J.}~\bibnamefont {Lantz}},\ and\ \bibinfo {author} {\bibfnamefont
  {G.}~\bibnamefont {Wendin}},\ }\href
  {https://doi.org/10.1103/PhysRevLett.90.087003} {\bibfield  {journal}
  {\bibinfo  {journal} {Phys. Rev. Lett.}\ }\textbf {\bibinfo {volume} {90}},\
  \bibinfo {pages} {087003} (\bibinfo {year} {2003})}\BibitemShut {NoStop}%
\bibitem [{\citenamefont {G\"ul}\ \emph
  {et~al.}(2014{\natexlab{b}})\citenamefont {G\"ul}, \citenamefont {G\"unel},
  \citenamefont {L\"uth}, \citenamefont {Rieger}, \citenamefont {Wenz},
  \citenamefont {Haas}, \citenamefont {Lepsa}, \citenamefont {Panaitov},
  \citenamefont {Gr\"utzmacher},\ and\ \citenamefont {Sch\"apers}}]{Guel2014a}%
  \BibitemOpen
  \bibfield  {author} {\bibinfo {author} {\bibfnamefont {O.}~\bibnamefont
  {G\"ul}}, \bibinfo {author} {\bibfnamefont {H.~Y.}\ \bibnamefont {G\"unel}},
  \bibinfo {author} {\bibfnamefont {H.}~\bibnamefont {L\"uth}}, \bibinfo
  {author} {\bibfnamefont {T.}~\bibnamefont {Rieger}}, \bibinfo {author}
  {\bibfnamefont {T.}~\bibnamefont {Wenz}}, \bibinfo {author} {\bibfnamefont
  {F.}~\bibnamefont {Haas}}, \bibinfo {author} {\bibfnamefont {M.}~\bibnamefont
  {Lepsa}}, \bibinfo {author} {\bibfnamefont {G.}~\bibnamefont {Panaitov}},
  \bibinfo {author} {\bibfnamefont {D.}~\bibnamefont {Gr\"utzmacher}},\ and\
  \bibinfo {author} {\bibfnamefont {T.}~\bibnamefont {Sch\"apers}},\ }\href
  {https://doi.org/10.1021/nl502598s} {\bibfield  {journal} {\bibinfo
  {journal} {Nano Letters}\ }\textbf {\bibinfo {volume} {14}},\ \bibinfo
  {pages} {6269–6274} (\bibinfo {year} {2014}{\natexlab{b}})},\ \bibinfo
  {note} {pMID: 25300066}\BibitemShut {NoStop}%
\bibitem [{\citenamefont {Zellekens}\ \emph {et~al.}(2024)\citenamefont
  {Zellekens}, \citenamefont {Deacon}, \citenamefont {Basaric}, \citenamefont
  {Juluri}, \citenamefont {Randle}, \citenamefont {Bennemann}, \citenamefont
  {Krause}, \citenamefont {Zimmermann}, \citenamefont {Sanchez}, \citenamefont
  {Gr{\"u}tzmacher}, \citenamefont {Pawlis}, \citenamefont {Ishibashi},\ and\
  \citenamefont {Sch{\"a}pers}}]{Zellekens2024}%
  \BibitemOpen
  \bibfield  {author} {\bibinfo {author} {\bibfnamefont {P.}~\bibnamefont
  {Zellekens}}, \bibinfo {author} {\bibfnamefont {R.~S.}\ \bibnamefont
  {Deacon}}, \bibinfo {author} {\bibfnamefont {F.}~\bibnamefont {Basaric}},
  \bibinfo {author} {\bibfnamefont {R.}~\bibnamefont {Juluri}}, \bibinfo
  {author} {\bibfnamefont {M.~D.}\ \bibnamefont {Randle}}, \bibinfo {author}
  {\bibfnamefont {B.}~\bibnamefont {Bennemann}}, \bibinfo {author}
  {\bibfnamefont {C.}~\bibnamefont {Krause}}, \bibinfo {author} {\bibfnamefont
  {E.}~\bibnamefont {Zimmermann}}, \bibinfo {author} {\bibfnamefont {A.~M.}\
  \bibnamefont {Sanchez}}, \bibinfo {author} {\bibfnamefont {D.}~\bibnamefont
  {Gr{\"u}tzmacher}}, \bibinfo {author} {\bibfnamefont {A.}~\bibnamefont
  {Pawlis}}, \bibinfo {author} {\bibfnamefont {K.}~\bibnamefont {Ishibashi}},\
  and\ \bibinfo {author} {\bibfnamefont {T.}~\bibnamefont {Sch{\"a}pers}},\
  }\bibfield  {journal} {\bibinfo  {journal} {arXiv preprint arXiv:2402.13880}\
  }\href {https://doi.org/10.48550/arXiv.2402.13880}
  {10.48550/arXiv.2402.13880} (\bibinfo {year} {2024})\BibitemShut {NoStop}%
\bibitem [{\citenamefont {Albrecht}\ \emph {et~al.}(2017)\citenamefont
  {Albrecht}, \citenamefont {Moers},\ and\ \citenamefont
  {Hermanns}}]{Albrecht2017}%
  \BibitemOpen
  \bibfield  {author} {\bibinfo {author} {\bibfnamefont {W.}~\bibnamefont
  {Albrecht}}, \bibinfo {author} {\bibfnamefont {J.}~\bibnamefont {Moers}},\
  and\ \bibinfo {author} {\bibfnamefont {B.}~\bibnamefont {Hermanns}},\ }\href
  {https://doi.org/10.17815/jlsrf-3-158} {\bibfield  {journal} {\bibinfo
  {journal} {Journal of large-scale research facilities JLSRF}\ }\textbf
  {\bibinfo {volume} {3}},\ \bibinfo {pages} {112} (\bibinfo {year}
  {2017})}\BibitemShut {NoStop}%
\end{thebibliography}%


\begin{thebibliography}{10}%
\makeatletter
\providecommand \@ifxundefined [1]{%
 \@ifx{#1\undefined}
}%
\providecommand \@ifnum [1]{%
 \ifnum #1\expandafter \@firstoftwo
 \else \expandafter \@secondoftwo
 \fi
}%
\providecommand \@ifx [1]{%
 \ifx #1\expandafter \@firstoftwo
 \else \expandafter \@secondoftwo
 \fi
}%
\providecommand \natexlab [1]{#1}%
\providecommand \enquote  [1]{``#1''}%
\providecommand \bibnamefont  [1]{#1}%
\providecommand \bibfnamefont [1]{#1}%
\providecommand \citenamefont [1]{#1}%
\providecommand \href@noop [0]{\@secondoftwo}%
\providecommand \href [0]{\begingroup \@sanitize@url \@href}%
\providecommand \@href[1]{\@@startlink{#1}\@@href}%
\providecommand \@@href[1]{\endgroup#1\@@endlink}%
\providecommand \@sanitize@url [0]{\catcode `\\12\catcode `\$12\catcode
  `\&12\catcode `\#12\catcode `\^12\catcode `\_12\catcode `\%12\relax}%
\providecommand \@@startlink[1]{}%
\providecommand \@@endlink[0]{}%
\providecommand \url  [0]{\begingroup\@sanitize@url \@url }%
\providecommand \@url [1]{\endgroup\@href {#1}{\urlprefix }}%
\providecommand \urlprefix  [0]{URL }%
\providecommand \Eprint [0]{\href }%
\providecommand \doibase [0]{https://doi.org/}%
\providecommand \selectlanguage [0]{\@gobble}%
\providecommand \bibinfo  [0]{\@secondoftwo}%
\providecommand \bibfield  [0]{\@secondoftwo}%
\providecommand \translation [1]{[#1]}%
\providecommand \BibitemOpen [0]{}%
\providecommand \bibitemStop [0]{}%
\providecommand \bibitemNoStop [0]{.\EOS\space}%
\providecommand \EOS [0]{\spacefactor3000\relax}%
\providecommand \BibitemShut  [1]{\csname bibitem#1\endcsname}%
\let\auto@bib@innerbib\@empty
\bibitem [{\citenamefont {Jansen}\ \emph {et~al.}(2020)\citenamefont {Jansen},
  \citenamefont {Perla}, \citenamefont {Kaladzhian}, \citenamefont {von~den
  Driesch}, \citenamefont {Janssen}, \citenamefont {Luysberg}, \citenamefont
  {Lepsa}, \citenamefont {Gr\"utzmacher},\ and\ \citenamefont
  {Pawlis}}]{Jansen2020}%
  \BibitemOpen
  \bibfield  {author} {\bibinfo {author} {\bibfnamefont {M.~M.}\ \bibnamefont
  {Jansen}}, \bibinfo {author} {\bibfnamefont {P.}~\bibnamefont {Perla}},
  \bibinfo {author} {\bibfnamefont {M.}~\bibnamefont {Kaladzhian}}, \bibinfo
  {author} {\bibfnamefont {N.}~\bibnamefont {von~den Driesch}}, \bibinfo
  {author} {\bibfnamefont {J.}~\bibnamefont {Janssen}}, \bibinfo {author}
  {\bibfnamefont {M.}~\bibnamefont {Luysberg}}, \bibinfo {author}
  {\bibfnamefont {M.~I.}\ \bibnamefont {Lepsa}}, \bibinfo {author}
  {\bibfnamefont {D.}~\bibnamefont {Gr\"utzmacher}},\ and\ \bibinfo {author}
  {\bibfnamefont {A.}~\bibnamefont {Pawlis}},\ }\href
  {https://doi.org/https://doi.org/10.1021/acsanm.0c02241} {\bibfield
  {journal} {\bibinfo  {journal} {ACS Applied Nano Materials}\ }\textbf
  {\bibinfo {volume} {3}},\ \bibinfo {pages} {11037} (\bibinfo {year}
  {2020})}\BibitemShut {NoStop}%
\bibitem [{\citenamefont {Webb}\ \emph {et~al.}(1986)\citenamefont {Webb},
  \citenamefont {Washburn}, \citenamefont {Umbach}, \citenamefont {Milliken},
  \citenamefont {Laibowitz},\ and\ \citenamefont {Benoit}}]{Webb86}%
  \BibitemOpen
  \bibfield  {author} {\bibinfo {author} {\bibfnamefont {R.}~\bibnamefont
  {Webb}}, \bibinfo {author} {\bibfnamefont {S.}~\bibnamefont {Washburn}},
  \bibinfo {author} {\bibfnamefont {C.}~\bibnamefont {Umbach}}, \bibinfo
  {author} {\bibfnamefont {F.}~\bibnamefont {Milliken}}, \bibinfo {author}
  {\bibfnamefont {R.}~\bibnamefont {Laibowitz}},\ and\ \bibinfo {author}
  {\bibfnamefont {A.}~\bibnamefont {Benoit}},\ }\href
  {https://doi.org/10.1016/0378-4371(86)90218-9} {\bibfield  {journal}
  {\bibinfo  {journal} {Physica A: Statistical Mechanics and its Applications}\
  }\textbf {\bibinfo {volume} {140}},\ \bibinfo {pages} {1039} (\bibinfo {year}
  {1986})}\BibitemShut {NoStop}%
\bibitem [{\citenamefont {Seelig}\ and\ \citenamefont
  {B\"uttiker}(2001)}]{Seelig2001}%
  \BibitemOpen
  \bibfield  {author} {\bibinfo {author} {\bibfnamefont {G.}~\bibnamefont
  {Seelig}}\ and\ \bibinfo {author} {\bibfnamefont {M.}~\bibnamefont
  {B\"uttiker}},\ }\href {https://doi.org/10.1103/PhysRevB.64.245313}
  {\bibfield  {journal} {\bibinfo  {journal} {Phys. Rev. B}\ }\textbf {\bibinfo
  {volume} {64}},\ \bibinfo {pages} {245313} (\bibinfo {year}
  {2001})}\BibitemShut {NoStop}%
\bibitem [{\citenamefont {Rosdahl}\ \emph {et~al.}(2014)\citenamefont
  {Rosdahl}, \citenamefont {Manolescu},\ and\ \citenamefont
  {Gudmundsson}}]{Rosdahl2014}%
  \BibitemOpen
  \bibfield  {author} {\bibinfo {author} {\bibfnamefont {T.~O.}\ \bibnamefont
  {Rosdahl}}, \bibinfo {author} {\bibfnamefont {A.}~\bibnamefont {Manolescu}},\
  and\ \bibinfo {author} {\bibfnamefont {V.}~\bibnamefont {Gudmundsson}},\
  }\href {https://doi.org/10.1103/PhysRevB.90.035421} {\bibfield  {journal}
  {\bibinfo  {journal} {Phys. Rev. B}\ }\textbf {\bibinfo {volume} {90}},\
  \bibinfo {pages} {035421} (\bibinfo {year} {2014})}\BibitemShut {NoStop}%
\bibitem [{\citenamefont {Sitek}\ \emph {et~al.}(2015)\citenamefont {Sitek},
  \citenamefont {Serra}, \citenamefont {Gudmundsson},\ and\ \citenamefont
  {Manolescu}}]{Sitek2015}%
  \BibitemOpen
  \bibfield  {author} {\bibinfo {author} {\bibfnamefont {A.}~\bibnamefont
  {Sitek}}, \bibinfo {author} {\bibfnamefont {L.}~\bibnamefont {Serra}},
  \bibinfo {author} {\bibfnamefont {V.}~\bibnamefont {Gudmundsson}},\ and\
  \bibinfo {author} {\bibfnamefont {A.}~\bibnamefont {Manolescu}},\ }\href
  {https://doi.org/10.1103/PhysRevB.91.235429} {\bibfield  {journal} {\bibinfo
  {journal} {Phys. Rev. B}\ }\textbf {\bibinfo {volume} {91}},\ \bibinfo
  {pages} {235429} (\bibinfo {year} {2015})}\BibitemShut {NoStop}%
\bibitem [{\citenamefont {Manolescu}\ \emph {et~al.}(2017)\citenamefont
  {Manolescu}, \citenamefont {Sitek}, \citenamefont {Osca}, \citenamefont
  {Serra}, \citenamefont {Gudmundsson},\ and\ \citenamefont
  {Stanescu}}]{Manolescu2017}%
  \BibitemOpen
  \bibfield  {author} {\bibinfo {author} {\bibfnamefont {A.}~\bibnamefont
  {Manolescu}}, \bibinfo {author} {\bibfnamefont {A.}~\bibnamefont {Sitek}},
  \bibinfo {author} {\bibfnamefont {J.}~\bibnamefont {Osca}}, \bibinfo {author}
  {\bibfnamefont {L.~m.~c.}\ \bibnamefont {Serra}}, \bibinfo {author}
  {\bibfnamefont {V.}~\bibnamefont {Gudmundsson}},\ and\ \bibinfo {author}
  {\bibfnamefont {T.~D.}\ \bibnamefont {Stanescu}},\ }\href
  {https://doi.org/10.1103/PhysRevB.96.125435} {\bibfield  {journal} {\bibinfo
  {journal} {Phys. Rev. B}\ }\textbf {\bibinfo {volume} {96}},\ \bibinfo
  {pages} {125435} (\bibinfo {year} {2017})}\BibitemShut {NoStop}%
\bibitem [{\citenamefont {Urbaneja~Torres}\ \emph {et~al.}(2018)\citenamefont
  {Urbaneja~Torres}, \citenamefont {Sitek}, \citenamefont {Erlingsson},
  \citenamefont {Thorgilsson}, \citenamefont {Gudmundsson},\ and\ \citenamefont
  {Manolescu}}]{Urbaneja2018}%
  \BibitemOpen
  \bibfield  {author} {\bibinfo {author} {\bibfnamefont {M.}~\bibnamefont
  {Urbaneja~Torres}}, \bibinfo {author} {\bibfnamefont {A.}~\bibnamefont
  {Sitek}}, \bibinfo {author} {\bibfnamefont {S.~I.}\ \bibnamefont
  {Erlingsson}}, \bibinfo {author} {\bibfnamefont {G.}~\bibnamefont
  {Thorgilsson}}, \bibinfo {author} {\bibfnamefont {V.}~\bibnamefont
  {Gudmundsson}},\ and\ \bibinfo {author} {\bibfnamefont {A.}~\bibnamefont
  {Manolescu}},\ }\href {https://doi.org/10.1103/PhysRevB.98.085419} {\bibfield
   {journal} {\bibinfo  {journal} {Phys. Rev. B}\ }\textbf {\bibinfo {volume}
  {98}},\ \bibinfo {pages} {085419} (\bibinfo {year} {2018})}\BibitemShut
  {NoStop}%
\bibitem [{\citenamefont {Doniach}\ and\ \citenamefont
  {Sondheimer}(1998)}]{Doniach1998}%
  \BibitemOpen
  \bibfield  {author} {\bibinfo {author} {\bibfnamefont {S.}~\bibnamefont
  {Doniach}}\ and\ \bibinfo {author} {\bibfnamefont {E.~H.}\ \bibnamefont
  {Sondheimer}},\ }\href@noop {} {\emph {\bibinfo {title} {Green's Functions
  for Solid State Physicists}}}\ (\bibinfo  {publisher} {Imperial College
  Press},\ \bibinfo {address} {London},\ \bibinfo {year} {1998})\BibitemShut
  {NoStop}%
\bibitem [{\citenamefont {Groth}\ \emph {et~al.}(2014)\citenamefont {Groth},
  \citenamefont {Wimmer}, \citenamefont {Akhmerov},\ and\ \citenamefont
  {Waintal}}]{Groth2014}%
  \BibitemOpen
  \bibfield  {author} {\bibinfo {author} {\bibfnamefont {C.~W.}\ \bibnamefont
  {Groth}}, \bibinfo {author} {\bibfnamefont {M.}~\bibnamefont {Wimmer}},
  \bibinfo {author} {\bibfnamefont {A.~R.}\ \bibnamefont {Akhmerov}},\ and\
  \bibinfo {author} {\bibfnamefont {X.}~\bibnamefont {Waintal}},\ }\href
  {https://doi.org/10.1088/1367-2630/16/6/063065} {\bibfield  {journal}
  {\bibinfo  {journal} {New Journal of Physics}\ }\textbf {\bibinfo {volume}
  {16}},\ \bibinfo {pages} {063065} (\bibinfo {year} {2014})}\BibitemShut
  {NoStop}%
\bibitem [{\citenamefont {Bommer}\ \emph {et~al.}(2019)\citenamefont {Bommer},
  \citenamefont {Zhang}, \citenamefont {G\"ul}, \citenamefont {Nijholt},
  \citenamefont {Wimmer}, \citenamefont {Rybakov}, \citenamefont {Garaud},
  \citenamefont {Rodic}, \citenamefont {Babaev}, \citenamefont {Troyer},
  \citenamefont {Car}, \citenamefont {Plissard}, \citenamefont {Bakkers},
  \citenamefont {Watanabe}, \citenamefont {Taniguchi},\ and\ \citenamefont
  {Kouwenhoven}}]{Bommer2019}%
  \BibitemOpen
  \bibfield  {author} {\bibinfo {author} {\bibfnamefont {J.~D.~S.}\
  \bibnamefont {Bommer}}, \bibinfo {author} {\bibfnamefont {H.}~\bibnamefont
  {Zhang}}, \bibinfo {author} {\bibfnamefont {O.}~\bibnamefont {G\"ul}},
  \bibinfo {author} {\bibfnamefont {B.}~\bibnamefont {Nijholt}}, \bibinfo
  {author} {\bibfnamefont {M.}~\bibnamefont {Wimmer}}, \bibinfo {author}
  {\bibfnamefont {F.~N.}\ \bibnamefont {Rybakov}}, \bibinfo {author}
  {\bibfnamefont {J.}~\bibnamefont {Garaud}}, \bibinfo {author} {\bibfnamefont
  {D.}~\bibnamefont {Rodic}}, \bibinfo {author} {\bibfnamefont
  {E.}~\bibnamefont {Babaev}}, \bibinfo {author} {\bibfnamefont
  {M.}~\bibnamefont {Troyer}}, \bibinfo {author} {\bibfnamefont
  {D.}~\bibnamefont {Car}}, \bibinfo {author} {\bibfnamefont {S.~R.}\
  \bibnamefont {Plissard}}, \bibinfo {author} {\bibfnamefont {E.~P. A.~M.}\
  \bibnamefont {Bakkers}}, \bibinfo {author} {\bibfnamefont {K.}~\bibnamefont
  {Watanabe}}, \bibinfo {author} {\bibfnamefont {T.}~\bibnamefont
  {Taniguchi}},\ and\ \bibinfo {author} {\bibfnamefont {L.~P.}\ \bibnamefont
  {Kouwenhoven}},\ }\href {https://doi.org/10.1103/PhysRevLett.122.187702}
  {\bibfield  {journal} {\bibinfo  {journal} {Phys. Rev. Lett.}\ }\textbf
  {\bibinfo {volume} {122}},\ \bibinfo {pages} {187702} (\bibinfo {year}
  {2019})}\BibitemShut {NoStop}%
\end{thebibliography}%

\newpage\null\newpage
\putbib[bu2.bbl] 
\end{bibunit}

\end{document}